\documentstyle[emulateapj,onecolfloat]{article}

\begin{document}
\twocolumn[

\title{The luminosities, sizes and velocity dispersions of 
       Brightest Cluster Galaxies: 
       Implications for formation history} 
\author{Mariangela Bernardi$^{1}$, Joseph B. Hyde$^{1}$, 
        Ravi K. Sheth$^{1}$, Chris J. Miller$^2$, and 
        Robert C. Nichol$^{3}$ }

\begin{abstract}
The size-luminosity relation of early-type Brightest Cluster 
Galaxies (BCGs),  $R_e\propto L^{0.88}$, 
is steeper than that for the bulk of the early-type galaxy 
population, for which $R_e\propto L^{0.68}$.  
This is true if quantities derived from either deVaucouleur or 
Sersic fits to the surface brightness profiles are used.  
Contamination from an intra-cluster light component centered on 
the BCG, with similar parameters to what has been seen in some 
recent studies, is not able to account for this difference.  
In addition, although BCGs are hardly offset from the Fundamental 
Plane defined by the bulk of the early-type population, they show 
considerably smaller scatter.  The larger than expected sizes of 
BCGs, and the increased homogeneity, are qualitatively consistent 
with models which seek to explain the colors of the most massive 
galaxies by invoking dry dissipationless mergers, since dissipation 
tends to reduce the sizes of galaxies, and wet mergers which result 
in star formation would tend to increase the scatter in luminosity 
at fixed size and velocity dispersion.  
Furthermore, BCGs define the same $g-r$ color-magnitude relation 
as the bulk of the early-type population.  
If BCGs formed from dry mergers, then BCG progenitors must have 
been red for their magnitudes, suggesting that they hosted older 
stellar populations than typical for their luminosities.  
Our findings have two other consequences.  
First, the $R_e-L$ relation of the early-type galaxy population 
as a whole (i.e., normal plus BCG) exhibits some curvature:  
the most luminous galaxies tend to have larger sizes than expected 
from the $R_e\propto L^{0.68}$ scaling---some of this curvature 
must be a consequence of the fact that an increasing fraction of 
the most luminous galaxies are BCGs.  
The second consequence is suggested by the fact that, despite 
following a steeper size-luminosity relation, BCGs tend to define 
a tight relation between dynamical mass $R_e\sigma^2/G$ and luminosity.  
Although this relation is slightly different than that defined by 
the bulk of the population, the fact that their sizes are large 
for their luminosities suggests that their velocity dispersions 
are small.  We find that, indeed, BCGs define a shallower $\sigma-L$ 
relation than the bulk of the early-type galaxy population.  
This shallower relation suggests there may be curvature in the 
correlation between black hole mass and velocity dispersion; 
simple extrapolation of a single power law $M_\bullet-\sigma$ 
relation to large $\sigma$ will underestimate $M_\bullet$.  
\end{abstract}

]

\footnotetext[1] {Department of Physics and Astronomy,
                 University of Pennsylvania, Philadelphia, PA 19104}
\footnotetext[2] {Cerro-Tololo Inter-American Observatory, NOAO, Casilla 603,
                  La Serena, Chile}
\footnotetext[3] {Institute of Cosmology and Gravitation (ICG), Mercantile House, Hampshire Terrace, University of Portsmouth, Portsmouth, PO1 2EG, UK}

\section{Introduction}
Understanding why massive early-type galaxies are red and dead 
has proved to be difficult; the problem is to arrange for star 
formation to occur at higher redshift than the actual assembly 
of the stars into a single massive galaxy.  The most recent galaxy 
formation models arrange for this to happen by a combination of 
two processes:  dry mergers and AGN feedback (Hopkins et al. 2005; 
Croton et al. 2005; De Lucia et al. 2005; Robinson et al. 2006; 
Bower et al. 2006).  
The dry merger hypothesis assumes that the assembly of massive 
galaxies occurs by merging smaller progenitor systems of old 
stars without additional star formation (which would otherwise 
lead to bluer colors).  This happens either because the merging 
units were themselves gas poor, or because AGN feedback has prevented 
the hot gas reservoirs of the progenitors from cooling and forming 
stars after the merger.  Together, these processes allow massive 
galaxies to be built from smaller systems while still remaining red.  

The dry merger hypothesis is most necessary for the most massive 
galaxies.  These tend to be the brightest galaxies in clusters 
(BCGs).  Our primary focus will be the sizes of these BCGs, although 
we also study a number of other scaling relations.  This is because, 
in a `wet' merger, gas from the merging components is able to 
dissipate energy, cool, contract and eventually make new stars.  
Dry mergers on the other hand have no energy loss mechanism 
(other than dynamical friction); having suffered less dissipation, 
their stellar components are not as centrally contracted, so the 
optical sizes of dry merger products are expected to be larger than 
if the mergers were wet (e.g. Kormendy 1989; Capelato et al. 1995; 
Nipoti et al. 2003), although the sizes may also reflect the orbital 
parameters of the mergers which formed the BCG 
(e.g. Boylan-Kolchin et al. 2006).  

\begin{figure*}
 \centering
 \epsfxsize=0.7\hsize\epsffile{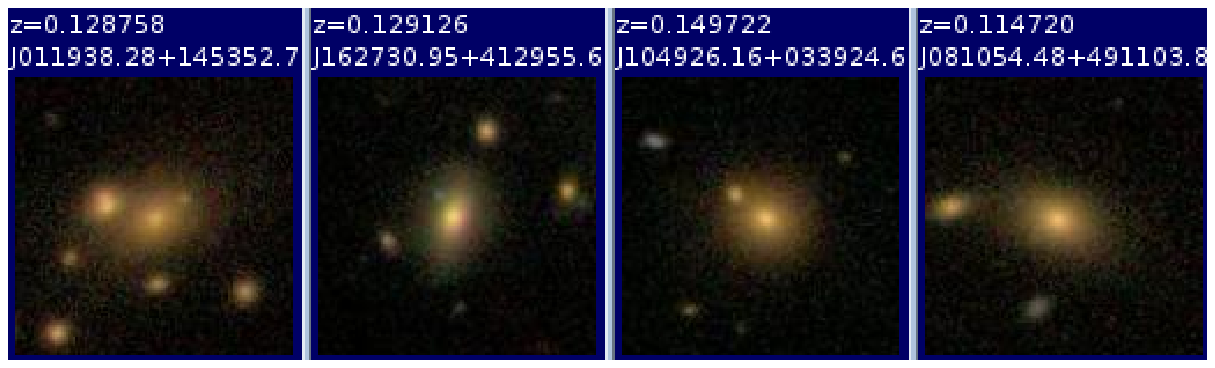}
 \epsfxsize=0.75\hsize\epsffile{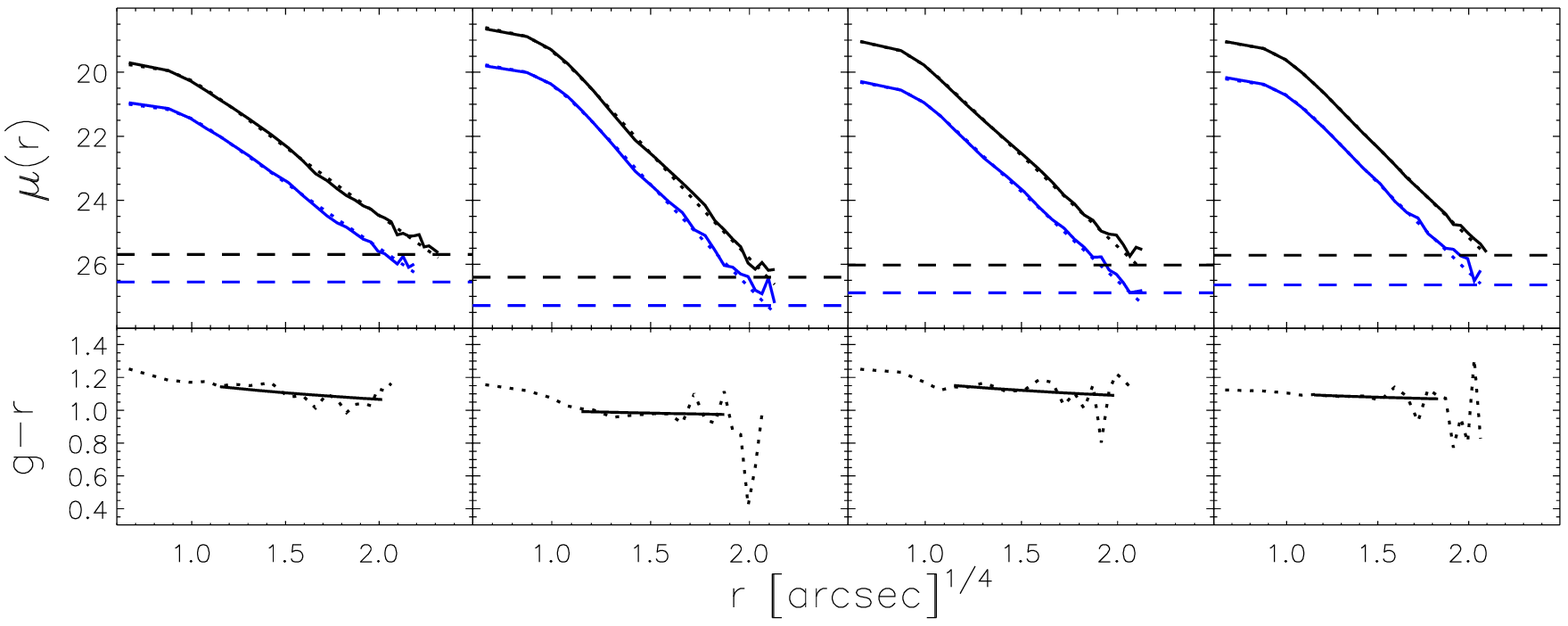}
 \caption{Images (top) and surface brightness profiles (bottom) 
          for a set of C4 BCGs which are well fit by a single 
          deVaucouleur profile.  
          The fit is good in both the $g$ and $r$ bands.  
          Horizontal dashed lines show one percent of the 
          sky brightness in the two bands.  
          Bottom half of bottom panels shows the associated 
          $g-r$ color gradient---the gradient is expected to 
          be weak if the formation was dominated by mergers.  }
 \label{puredeV}
\end{figure*}

The structural properties of BCGs have been the object of much 
previous work.  Compared to normal early type galaxies, BCGs have 
larger than expected radii for their surface brightnesses 
(Thuan \& Romanishin 1981; Hoessel et al. 1987; Schombert 1987, 1988; 
Oegerle \& Hoessel 1991), and smaller than expected velocity 
dispersions for their luminosities (Malumuth \& Kirshner 1981, 1985; 
Oegerle \& Hoessel 1991).  
Crawford et al. (1999) describe the results of an extensive study 
of the spectra of X-ray selected BCGs, and Brough et al. (2005) 
find that BCGs of X-ray luminous clusters tend to be have larger 
than expected sizes.  With the exception of Crawford et al., 
these studies were based on sample sizes of order 50 or smaller, 
so it is interesting to ask if the larger samples now available 
show the same trends.  

We will study the objects which are identified as BCGs in the C4 
cluster catalog of Miller et al. (2005).  
This catalog is complete out to $z=0.12$, contains about 700 BCGs, 
and will contain about 1500 BCGs by Fall 2006.  
Section~\ref{photo} describes the sample and provides a brief summary 
of our fits to the photometric properties of BCGs and normal 
early-type galaxies.  Details are provided in Hyde et al. (2006). 
These new fits were necessary, as the SDSS photometric reductions 
of objects in crowded fields suffer from sky subtraction problems 
(e.g. Hyde et al. 2006, Lauer et al. 2006).  
In Section~\ref{scaling} we compare a variety of BCG scaling 
relations to those of the bulk of the early-type galaxy population.  
This section shows that BCGs define a steeper size-luminosity 
relation than does the bulk of the early-type galaxy population; 
that although BCGs have larger than expected sizes, they still define 
a tight relation between luminosity and dynamical mass $R\sigma^2/G$, 
and so the $\sigma-L$ relation of BCGs curves away from the power-law 
scaling defined by the bulk of the early-type galaxy population. 
It also shows the location of BCGs in the same Fundamental Plane 
defined by normal early-type galaxies, and shows that they define 
the same color-magnitude relation as normal early-type galaxies.  
Appendix~A presents a comparison with the color-magnitude relation 
predicted by the semi-analytic galaxy formation models of 
Croton et al. (2006) and Bower et al. (2006).  

Section~\ref{systematics} studies two systematic errors which might 
have caused the steeper $R_e-L$ relation we see, and concludes 
that they are probably not to blame.  
The first possibility is that the deVaucouleur fit is systematically 
worse for BCGs than it is for the bulk of the population.  
The second possibility is more subtle.  
Galaxy clusters contain a substantial population of stars which 
are not associated with a galaxy (e.g. Gonzalez et al. 2005; 
Zibetti et al. 2005).  So it is not unreasonable to ask if this 
intracluster light is contaminating our estimates of the sizes and 
luminosities of BCGs, and hence the inferred $R_e-L$ and $\sigma-L$ 
relations.  

At the very largest luminosities ($M_r<-23.5$), the $R_e-L$ 
relation of the early-type galaxy population (normal + BCG) appears 
to curve upwards toward larger sizes from the $R_e\propto L^{0.68}$ 
scaling.  If an increasing fraction of these objects are BCGs, then 
this curvature is not unexpected in view of our finding that BCGs 
follow a steeper size-luminosity relation.  Appendix~\ref{bigL} 
presents evidence that, at large luminosities, an increasing fraction 
of galaxies are indeed BCGs.  However, our results do not exclude 
the possibility that there is, in addition, intrinsic curvature in 
the $R_e-L$ relation of early-types which are not BCGs.

A final section summarizes our findings and discusses the 
implications for galaxy formation models, and for studies which 
seek to estimate the mass of a supermassive black hole from the 
stellar velocity dispersion of the bulge which surrounds it.  
Where necessary, we assume a flat cosmological model with 
$\Omega_0 = 0.3$, $\Lambda=1-\Omega_0$ 
and $H_0 = 70h_{70}$~km~s$^{-1}$Mpc$^{-1}$.  

\begin{figure*}
 \centering
 \epsfxsize=0.7\hsize\epsffile{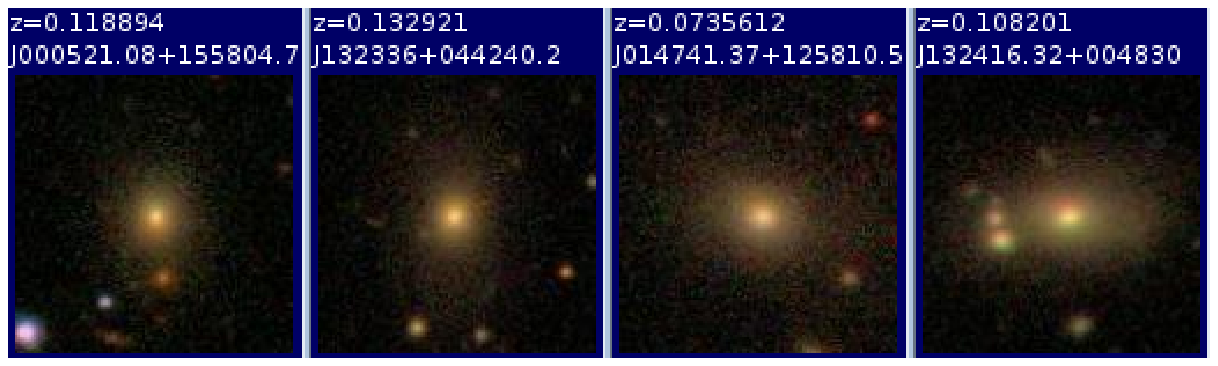}
 \epsfxsize=0.75\hsize\epsffile{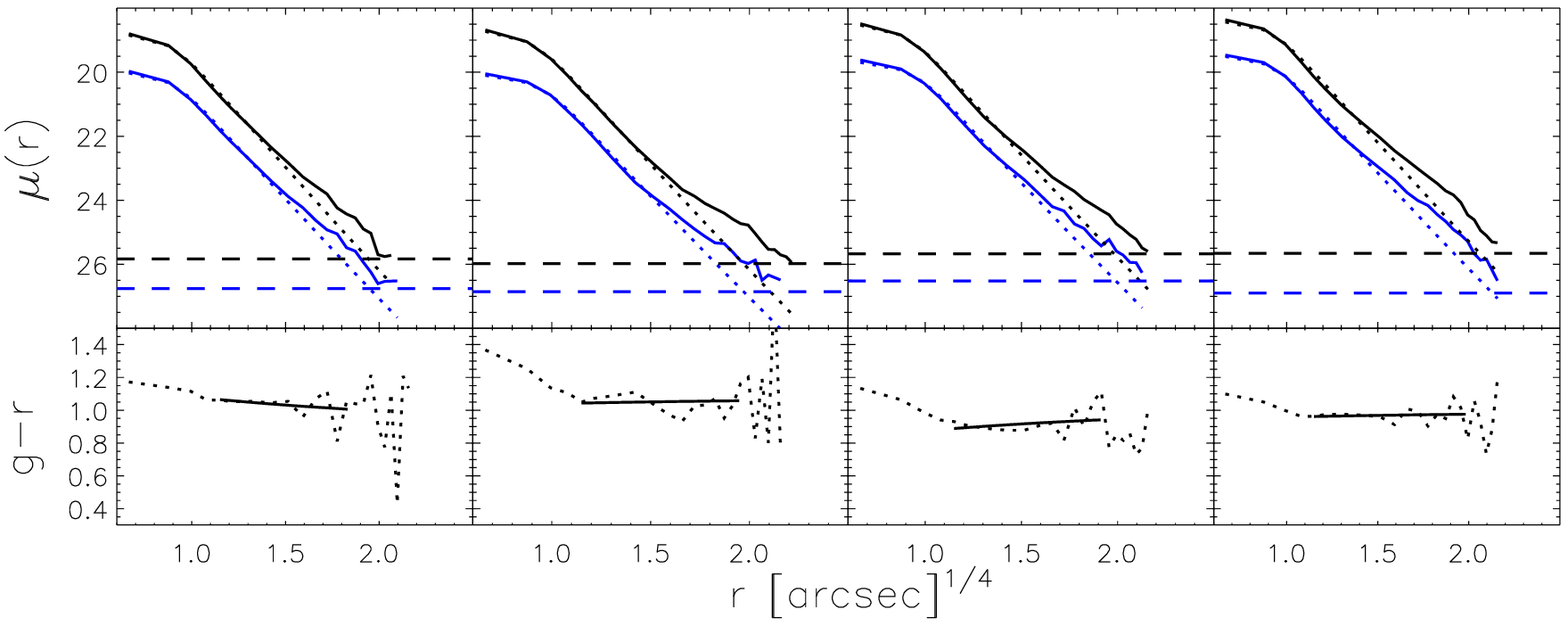}
 \caption{Images (top) and surface brightness profiles (bottom) 
          for a set of C4 BCGs which are not well fit by a deVaucouleur 
          profile.  The deviation is strongest at large radii (faint 
          surface brightness).  Although Sersic profiles provide 
          better fits, Gonzalez et al. (2006) show that the sum of 
          two deVaucouleur components generally provide an even 
          better description of most such profiles.}
 \label{twodeVs}
\end{figure*}

\section{Sample selection and fits to light profiles}\label{photo}
We have selected the objects identified as BCGs in the C4 cluster 
catalog of Miller et al. (2005).  This catalog was based on the 
Second Data Release of the SDSS: DR2.  
Of the 748 BCGs in the catalog, spectra are available for 580 
(fiber collisions mean some BCGs were missed).  
Some of these are duplicates (i.e. some clusters were listed 
more than once), and the number which are matched to a parent 
catalog based on SDSS DR4 is 403.  Of these, a visual inspection 
of the fields centred on C4 BCGs fainter than $M_r<-21$ often showed 
a brighter galaxy nearby, especially when the C4 BCG was a spiral.  
If we restrict attention to objects with $M_r<-21$, then 286 have 
SDSS reported velocity dispersions, and have at least 10 other 
luminous galaxies ($M_r<-21$) within $1h^{-1}$Mpc.  
The SDSS does not estimate velocity dispersions from spectra 
which have strong emission lines, so our requirement that the 
SDSS reports a velocity dispersion means that the objects we select 
are almost certainly not AGN.  Removing the final cut on `richness' 
increases the sample of BCGs to 324, but does not change the results 
which follow.
  
Visual inspection of the SDSS images of these objects shows that 
25\% actually have spiral arms; the absence of emission lines in 
the spectra of these objects is presumably because the SDSS fiber, 
with a radius of 1.5 arcsec, only sees the inner bulge.  
In what follows, we will be careful to distinguish the scaling 
relations of the $\sim 215$ genuine early-type BCGs from the rest.  

It happens that BCGs are a class of object for which the SDSS 
photometric reductions are particularly unreliable:  the reported 
magnitudes are in error by far more than the quoted 0.02~mags, 
with some discrepancies as large as 1~mag (Hyde et al. 2006).  
This is due to a combination of factors:  because BCGs are often 
in crowded fields, the SDSS reductions tend to overestimate the 
sky level.  
This results in underestimates of the BCG magnitude and 
half-light radius, which are particularly severe for BCGs with 
large half-light radii.  
In addition, the surface brightness distributions of BCGs do 
not always follow pure single component deVaucouleurs profiles, 
which the SDSS reductions assume.  For these reasons, we have 
performed our own photometric reductions, and examined the image 
of each BCG by eye.  A more detailed discussion of the various 
issues involved is provided by Hyde et al. (2006), whose results 
we use below.  

Because we will be interested in comparing the properties of 
BCGs to those of other early-type galaxies, we have also 
reanalyzed the photometric properties of the sample of $9000$ 
SDSS early-type galaxies defined in Bernardi et al. (2003a). 
It is these parameters (i.e. those based on Hyde et al. 2006), 
rather than those output from the SDSS pipeline, which we use in 
what follows.  On average, the Hyde et al. magnitudes are 0.1~mags 
brighter than those reported by the SDSS DR4 pipeline, and the 
sizes are about 10\% smaller.  In crowded fields, and especially 
for bright objects, Hyde et al. report both brighter magnitudes 
and larger sizes than does the SDSS.  
When comparing the scaling relations of these objects with those 
for the C4 BCGs, we restrict the redshift range to that over which 
the C4 catalog is complete ($z<0.12$; except for the Fundamental 
Plane, this restriction makes little difference).  

The most important finding of this reanalysis of BCG photometry 
is that there appear to be two classes of early-type BCG light 
profiles.  Examples of these two types are shown in 
Figures~\ref{puredeV} and~\ref{twodeVs}.  The first type is well 
fit by a single component deVaucouleurs profile, whereas the second 
is not.  This second type of profile is similar to that associated 
with cD galaxies.  Section~\ref{systematics} discusses these two 
profile types in more detail.
Meantime, we will be careful to treat these two populations 
separately in what follows.  

\begin{figure}
 \centering
 \epsfxsize=\hsize\epsffile{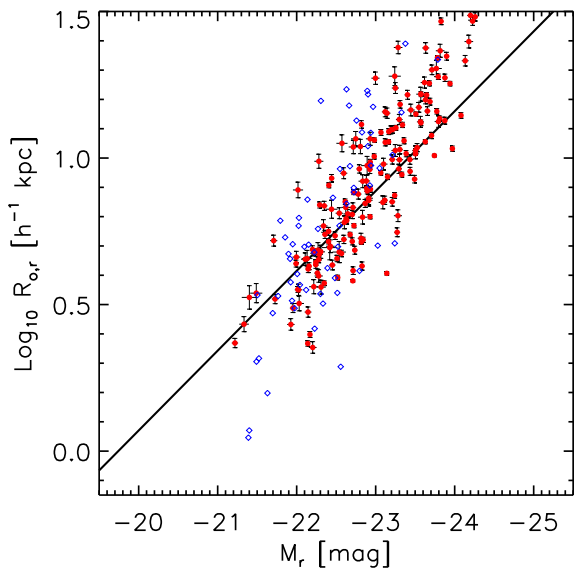}
 \caption{C4 BCGs define a steeper size-luminosity relation (symbols) 
          than does the bulk of the early-type galaxy population 
          (solid line).  Open diamonds and filled circles represent 
          BCGs with spiral and E/S0 morphologies, respectively.  
          In what follows, we only consider the E/S0 objects.}
 \label{morphLRe}
\end{figure}

\begin{figure}
 \centering
 \epsfxsize=\hsize\epsffile{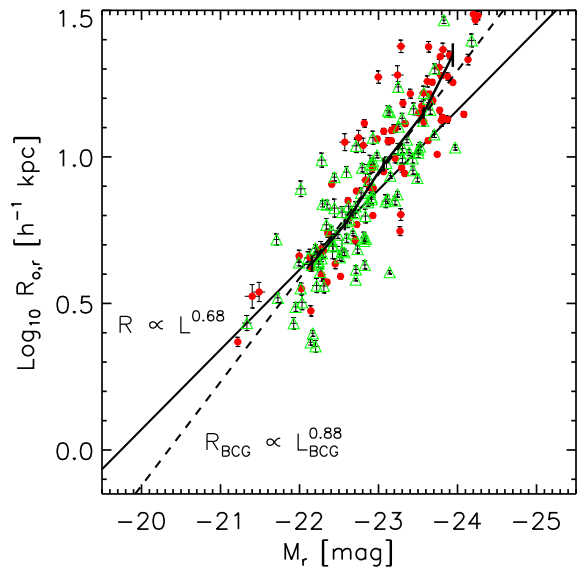}
 \caption{Early-type C4 BCGs define a steeper size-luminosity 
          relation (symbols) than does the bulk of the early-type 
          galaxy population (solid line).  Filled circles represent 
          objects which are well-described by a deVaucouleurs 
          profile, and open triangles represent objects which are not.  
          Dashed line shows a fit to the $R_e-L$ relation using all 
          these BCGs (i.e. both filled and open symbols), and jagged 
          solid line shows the median BCG size as a function of luminosity.}
 \label{LRe}
\end{figure}

\section{BCG scaling relations}\label{scaling}
\subsection{The size-luminosity relation of BCGs}\label{ReL}
Figure~\ref{morphLRe} shows the correlation between luminosity 
and half-light radius for the C4 BCGs:  diamonds represent objects 
with spiral morphologies, and filled circles represent E/S0s.  
Solid line shows the $R_e-L$ relation defined by the bulk of the 
early-type galaxy population---BCGs clearly define a steeper 
relation.  (The steeper relation is also seen if we use the 
original SDSS photometric reductions which we believe to be 
incorrect---see bottom right panel of Figure~\ref{allLRe}.)  
In what follows, we focus on the early-type BCGs, but note that 
the spiral BCGs define an $R_e-L$ relation that is at least as 
steep.   

Figure~\ref{LRe} shows the $R_e-L$ relation for the early-type 
BCGs, now subdivided by whether (filled circles) or not (open 
triangles) the surface brightness profile is well fit by a single 
deVaucouleurs profile.  Jagged solid line shows the median BCG 
size (whether or not it is well-fit by a deVaucouleur profile) 
in a few bins in luminosity, and dashed line shows a power-law 
fit to the size-luminosity relation for these BCGs:
\begin{equation}
 \Bigl\langle \log_{10}R_e|M_r\Bigr\rangle = -0.885\, (M_r+21)/2.5 + 0.230.
 \label{lreBCG}
\end{equation} 
This fit should be compared to the straight solid line which 
shows the scaling that provides a good description of the bulk 
of the early-type galaxy population:
\begin{equation}
 \Bigl\langle \log_{10}R_e|M_r\Bigr\rangle = -0.681\,(M_r+21)/2.5 + 0.343. 
 \label{lreALL}
\end{equation}
Clearly, BCGs define a steeper relation---they have larger sizes 
than do normal early-type galaxies of similar luminosity.  
This trend is consistent with the hypothesis that BCGs have 
formation histories with less dissipation than the bulk of 
the early-type population.  

The fits above are based on the photometric reductions described 
by Hyde et al. (2006).  Using the SDSS photometry (deVaucouleur magnitudes 
and sizes) instead gives 
\begin{displaymath}
 \Bigl\langle \log_{10}R_e|M_r\Bigr\rangle = -0.764\, (M_r+21)/2.5 + 0.395
\end{displaymath}
for the BCGs and  
\begin{displaymath}
 \Bigl\langle \log_{10}R_e|M_r\Bigr\rangle = -0.623\,(M_r+21)/2.5 - 0.402
\end{displaymath}
for the total early-type galaxy population.

\subsection{The mass-luminosity relation}\label{ML}
\begin{figure}
 \centering
 \epsfxsize=\hsize\epsffile{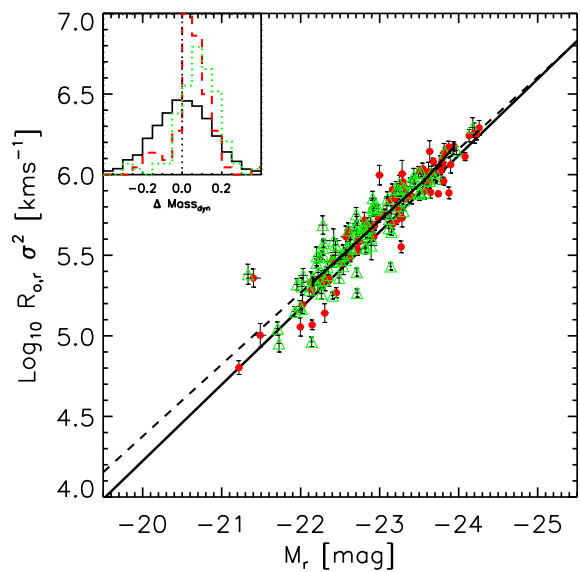}
 \caption{Correlation between dynamical mass $R_e\sigma^2$ and 
          luminosity in the bulk of the early-type population 
          (solid line) and in the BCG sample (dashed line); 
          filled circles and open triangles represent BCGs which 
          are and are not well-fit by a deVaucouleurs profile.  
          Solid jagged line shows the median value as a function 
          of $L$.  BCGs define a slightly shifted and even tighter 
          correlation between dynamical mass $R_e\sigma^2$ and 
          luminosity than does the bulk of the early type galaxy 
          population (inset).  }
 \label{MLbcg}
\end{figure}

The larger than expected sizes of BCGs have interesting consequences.  
If these objects are virialized, then one expects a relatively 
tight correlation between luminosity and the dynamical mass 
$\propto R_e\sigma^2/G$.  
Figure~\ref{MLbcg} shows that this correlation is indeed rather 
tight---the inset shows that BCGs define a tighter relation than 
does the bulk of the early-type galaxy population.  This smaller 
scatter is also consistent with dry-merger formation histories; 
one might have expected `wet' mergers to result in star formation, 
which would increase the scatter in luminosities at fixed dynamical 
mass.  

\begin{figure}
 \centering
 \epsfxsize=\hsize\epsffile{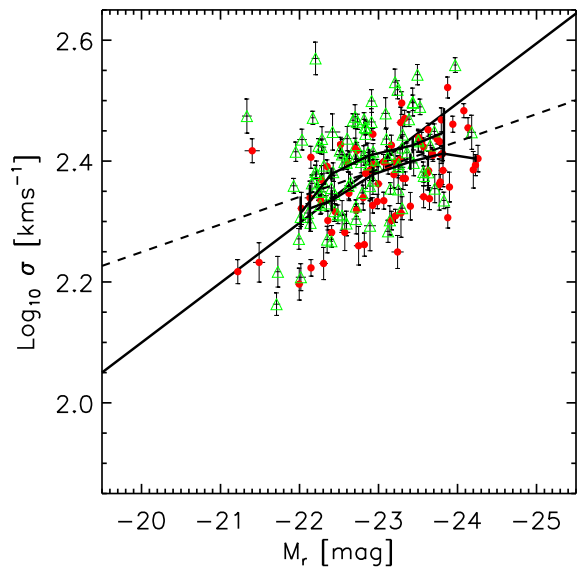}
 \caption{The $\sigma-L$ relation of C4 BCGs (filled and open 
          symbols represent objects which are and are not well 
          described by a deVaucouleurs profile) appears to 
          flatten at high luminosities.  
          This flattening may be related to studies of curvature 
          in the $M_\bullet-\sigma$ relation.  
          Solid and dashed lines show linear fits to this relation 
          for the early-type galaxy population and for this BCG 
          sample. Solid jagged lines show the median value as a function 
          of $L$ for objects which are (lower line) and are not 
          well-fit (upper line) by a single deVaucouleur profile. 
             }
 \label{LSigma}
\end{figure}

Figure~\ref{MLbcg} also shows that BCGs define a slightly 
different correlation between luminosity and $R_e\sigma^2$ than 
does the total early-type galaxy population.  The actual relations 
are
\begin{equation}
 \Bigl\langle \log_{10}R_e\sigma^2|M_r\Bigr\rangle 
 = -1.114\,(M_r+21)/2.5 + 4.822
\end{equation}
with a scatter around the relation of 0.11~dex for the BCGs and  
\begin{equation}
 \Bigl\langle \log_{10}R_e\sigma^2|M_r\Bigr\rangle 
 = -1.185\, (M_r+21)/2.5 - 4.697
\end{equation}
with a scatter around the relation of 0.34~dex for the total early-type galaxy population.  

\subsection{The velocity dispersion-luminosity relation}\label{VL}
If the $R_e-L$ relation for BCGs is steeper than for the bulk 
of the early-type galaxy population, then the tight correlation 
between $L$ and dynamical mass shown in Figure~\ref{MLbcg} 
suggests that the $\sigma-L$ relation for BCGs will be shallower.  
Figure~\ref{LSigma} shows that this does appear to be the 
case.  Once again, the solid line shows the relation for the 
full sample, and the dashed line shows a fit to the BCGs (filled 
and open symbols).  
A running average over small bins in luminosity shows evidence 
for curvature in the BCG $\sigma-L$ relation, in the sense that 
velocity dispersion increases only weakly with luminosity once 
$\sigma\ge$250~km/s.  However, notice that objects which are not 
well-fit by a single deVaucouleur profile appear to have slightly 
larger velocity dispersions.  

\begin{figure}
 \centering
 \epsfxsize=0.95\hsize\epsffile{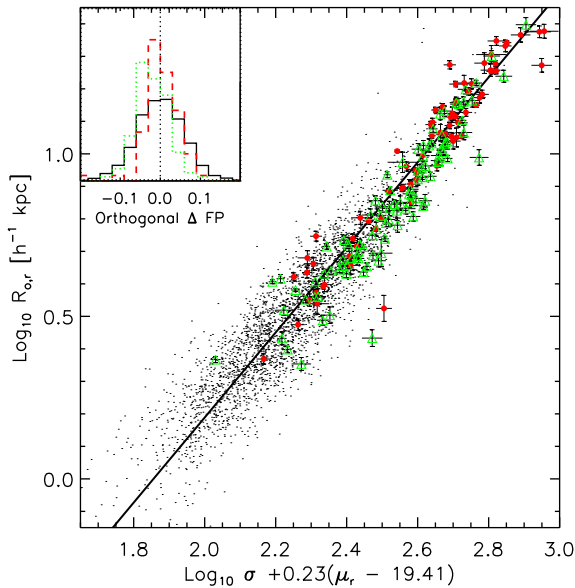}
 \caption{Location of C4 BCGs (filled and open symbols as in 
          previous figures) with respect 
          to the Fundamental Plane defined by the bulk of the 
          early-type galaxy population (dots).  Inset shows 
          that BCGs are distributed more tightly around this FP 
          than the bulk of the early-type galaxy population:  
          the rms scatter is 0.04~dex compared to 0.058~dex. } 
 \label{fpbcg}
\end{figure}

\begin{figure*}
 \centering
 \epsfxsize=0.45\hsize\epsffile{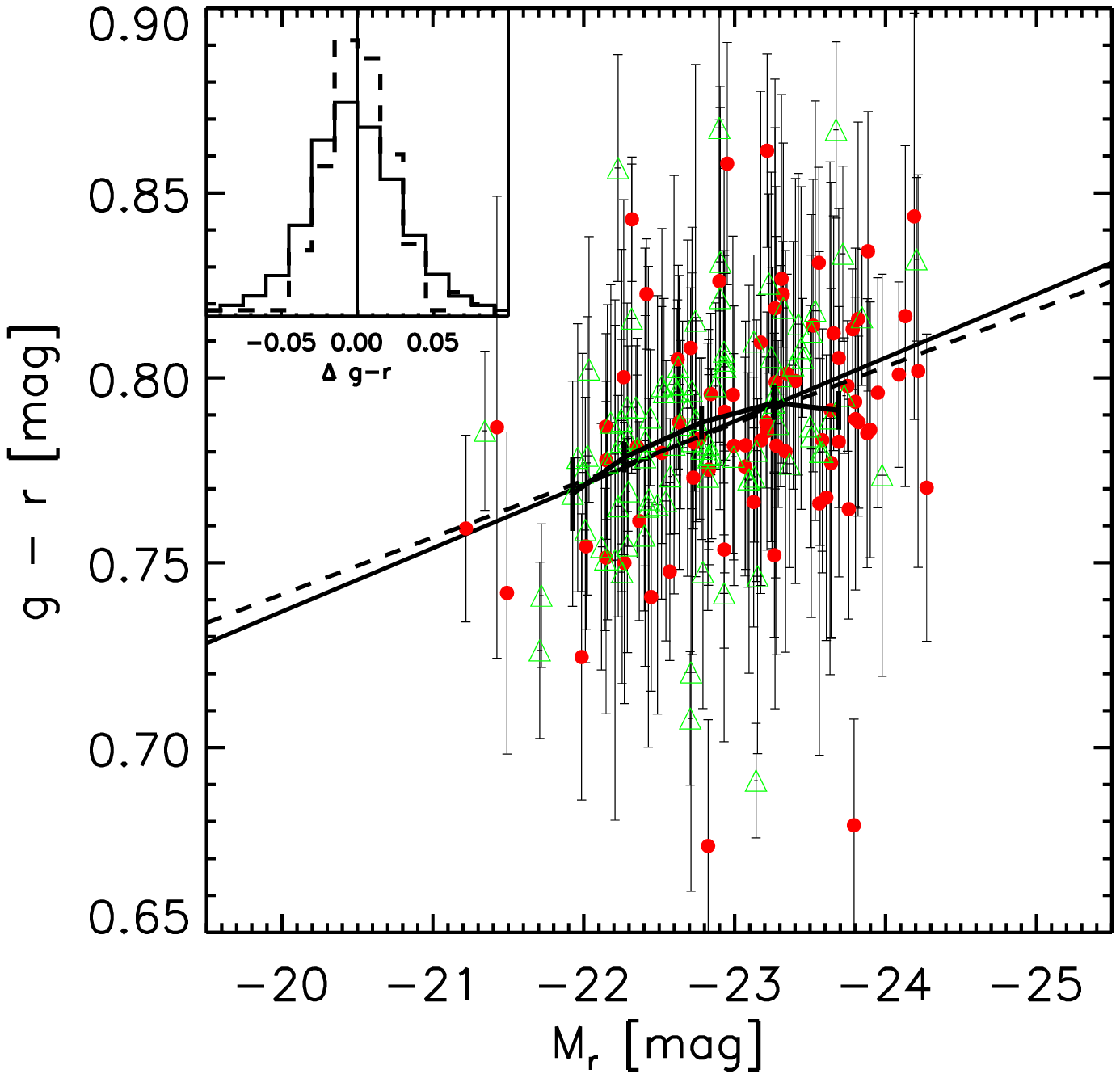}
 \epsfxsize=0.45\hsize\epsffile{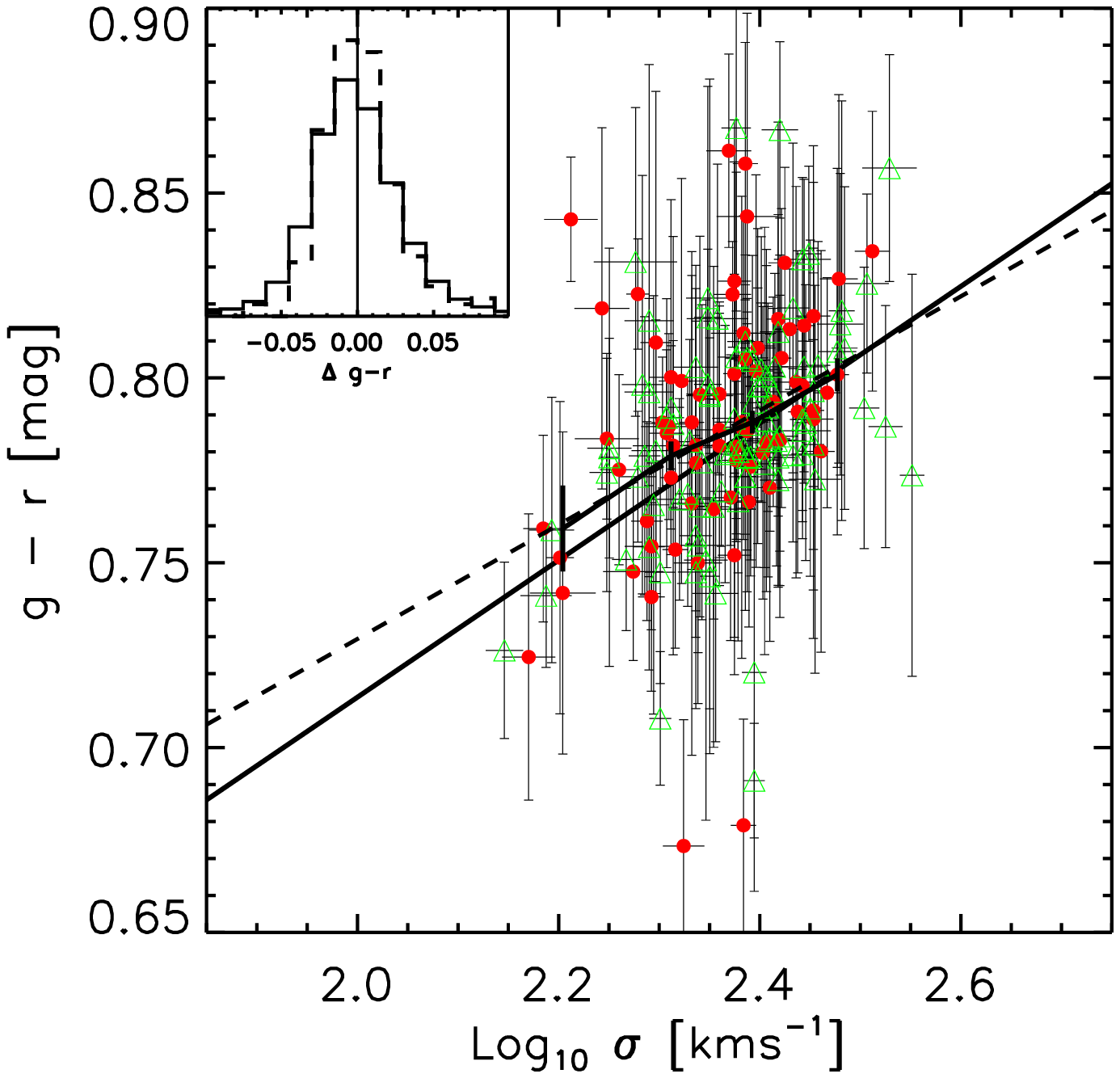}
 \caption{Location of C4 BCGs (filled and open symbols represent 
          objects which are and are not well-fit by a deVaucouleurs 
          profile) with respect to the color-magnitude (left) and 
          color-$\sigma$ (right) relations defined by the bulk of 
          the early-type galaxy population (solid line).  
          Dashed line shows a linear fit of color as 
          a function of absolute magnitude (or velocity dispersion), 
          and jagged line with 
          error bars shows the median color in a few bins in luminosity
          (or velocity dispersion).  
          Inset shows that BCGs are not offset from the relation 
          defined by the bulk of the population, though the relation 
          they define is slightly tighter.  }
 \label{cmbcg}
 \label{csbcg}
\end{figure*}

\subsection{The Fundamental Plane}\label{FP}
Numerical simulations of a dry merger scenario suggest that the 
final object will remain on the Fundamental Plane defined by 
the bulk of the early-type population (e.g. Boylan-Kolchin et al. 2006).  
Figure~\ref{fpbcg} shows the location of the C4 BCGs in the 
Fundamental Plane defined by the bulk of the early-type galaxy 
population:  
\begin{equation}
 \log_{10}R_e = 1.307\, \sigma + 0.763\, \mu_r/2.5 - 8.345.
\end{equation}
This is the ``orthogonal'' Fundamental Plane and it differs from 
the result presented in Bernardi et al. (2003b) mainly because we 
have restricted our sample to $z < 0.12$ (the new photometry 
changed the zero-point but not the slope of the relation).

The BCGs which are well fit by a single deVaucouleur profile lie 
on this same Fundamental Plane, whereas the others are slightly 
offset ($-0.025$~dex) from it.  The inset shows that the scatter 
around this Plane is significantly reduced---the scatter orthogonal 
to the Plane decreases from 0.058~dex (solid line), to 0.046~dex 
(dotted line), to 0.04~dex (dashed-line) if only BCGs which are 
well fit by a deVaucouleur profile are used.  
This reduced scatter for BCGs is consistent with a dry merger 
formation history---even a small amount of star formation 
associated with a wet merger has the potential to change the 
surface brightnesses significantly, resulting in increased 
scatter about the Plane.

\subsection{The color-magnitude relation}\label{CM}


Early-type galaxies define a tight correlation between color and 
luminosity.  Figure~\ref{cmbcg} shows that BCGs follow the same 
correlation, though with slightly smaller scatter.  
The Appendix presents a comparison with the color-magnitude 
relation predicted by semi-analytic galaxy formation models.  
The models suggest that BCGs should lie slightly blueward of the 
relation defined by the bulk of the population---an effect which 
is not seen in Figure~\ref{cmbcg}.  

Whereas the luminosity of an object formed from a dry merger is 
expected to be the sum of the luminosities of the objects involved 
in the merger, the merger is not expected to change the color 
(there being no associated star formation).  
Therefore, if BCGs formed from dry mergers, then the fact that they 
define the {\em same} color-magnitude relation as the bulk of the 
population suggests that the progenitors of BCGs were red for their 
luminosities.  Objects which lie redward of the color-magnitude 
relation are expected to be older (Kodama et al. 1998; 
Bernardi et al. 2005).  Thus, Figure~\ref{cmbcg} is not inconsistent 
with the expectation that the oldest stars in a halo find their way 
to the bottom of the potential well, the location of which is 
often traced by the BCG.  

\begin{figure*}
 \centering
 \epsfxsize=0.32\hsize\epsffile{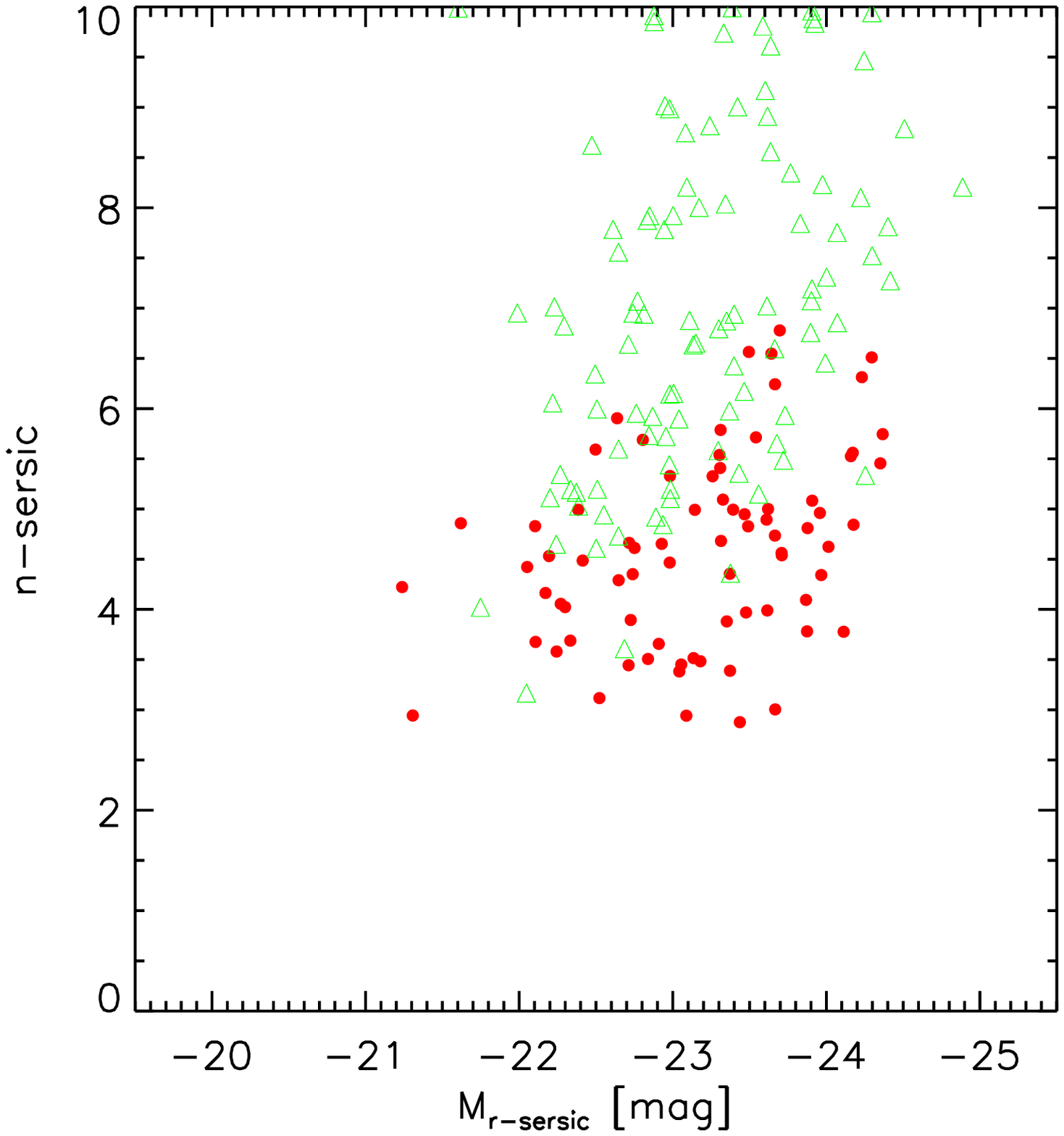}
 \epsfxsize=0.32\hsize\epsffile{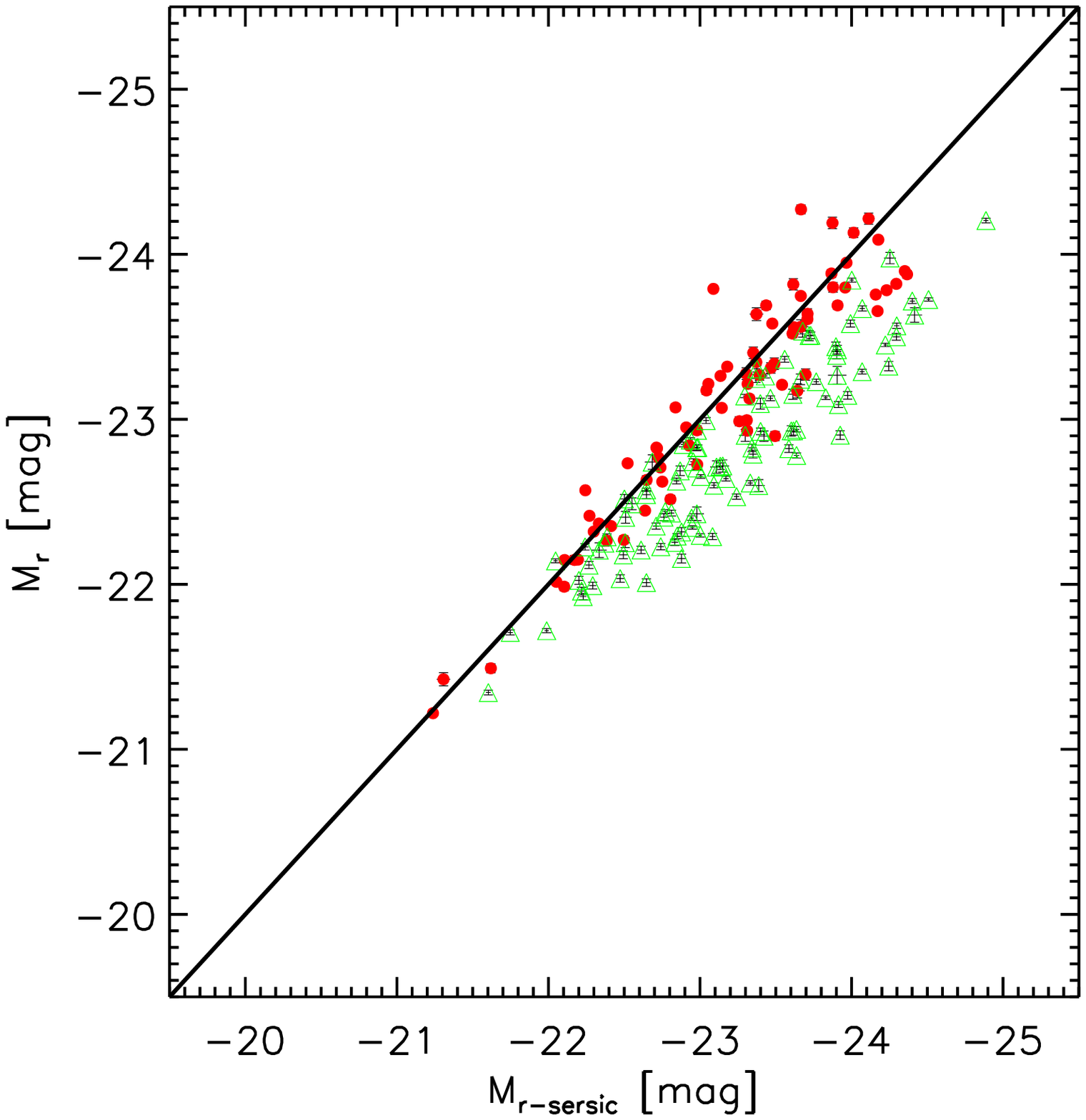}
 \epsfxsize=0.32\hsize\epsffile{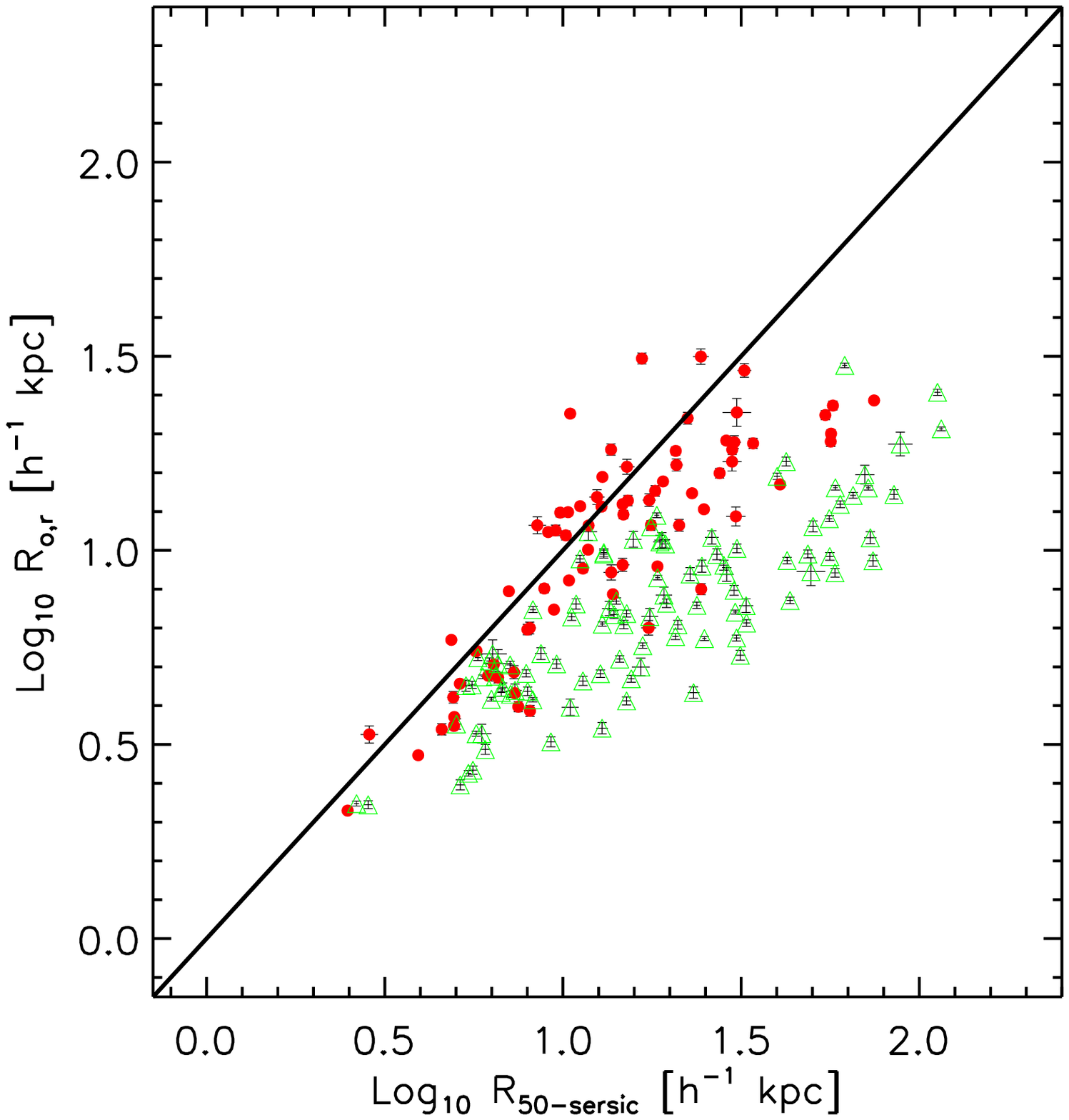}
 \caption{Left: Correlation between Sersic parameter $n$ and the 
          total luminosity obtained from fitting the Sersic 
          form to the surface brightness profiles.  Filled and 
          open symbols are the same as in previous figures.  
          The filled symbols tend to have $n=4$, 
          the value which represents a deVaucouleur profile, 
          whereas objects which show evidence for two components 
          tend to have large $n$.  
           When the deVaucouleur profile is a poor fit (open symbols), 
           the Sersic fits result in significantly larger total 
           luminosities (middle panel) and sizes (right).}
 \label{nLsersic}
 \label{devSersic}
\end{figure*}

Figure~\ref{csbcg} shows that BCGs also lie on approximately the same 
$\sigma$-color correlation as the bulk of the early-type population.  
This is consistent with a model in which the color is not altered by 
the merger, and the change to $\sigma$ is also weak (recall 
Figure~\ref{LSigma} suggests that dry mergers do not change 
$\sigma$ as much as they change $L$).  

The color-magnitude relation for the bulk of the early-type galaxy 
population is entirely due to the color-$\sigma$ and 
$\sigma$-magnitude correlations (Bernardi et al. 2005).  
The figures above suggest that this is not true for BCGs.
Specifically, the flatter $\sigma-L$ relation for BCGs, combined 
with the fact that the color-magnitude and color$-\sigma$ relations 
show little difference from the main sample implies that the 
correlation between color and magnitude for BCGs is substantially 
stronger than expected from the BCG color-$\sigma$ and 
$\sigma$-magnitude correlations.

\section{Tests of systematic effects}\label{systematics}
This section studies two possible sources of systematic trends 
and concludes that they do not cause the steeper $R_e-L$ relation 
we see in our early-type BCG sample.


\subsection{Sersic profile fits}\label{sersic}

\begin{figure}
 \centering
 \epsfxsize=\hsize\epsffile{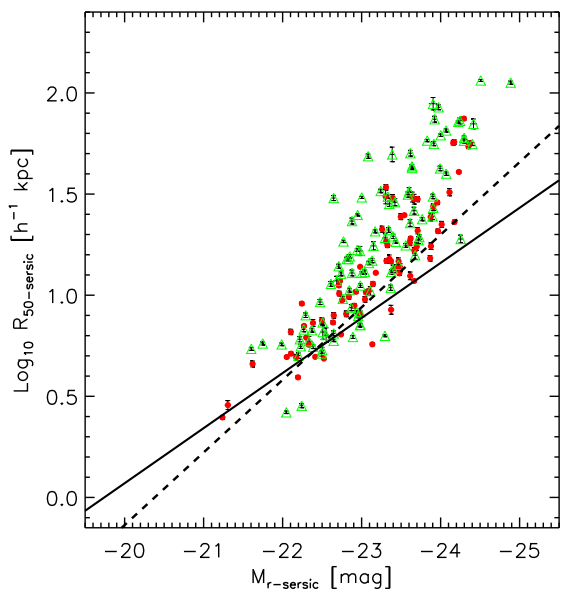}
 \caption{The $R_e-L$ relation obtained from fitting Sersic profiles 
          to the surface-brightness distribution (compare 
          Figure~\ref{LRe}).  
          Solid line shows the $R_e-L$ relation defined by the bulk 
          of the early-type galaxy population and dashed line shows 
          the steeper relation defined by the C4 BCGs when the fit 
          is restricted to $n=4$, a deVaucouleur profile; symbols 
          show that C4 BCGs define a steeper relation even when 
          Sersic parameters are used.  Filled symbols 
          represent objects which are well fit by a single component 
          deVaucouleur profile in the previous figures, and open 
          symbols show the other early-type BCGs.}
 \label{RLsersic}
\end{figure}

We have considered the possibility that the steeper $R_e-L$ 
relation we see arises because the deVaucouleur profile provides 
a systematically worse description of early-type surface brightness 
profiles at large $L$.  
Figures~\ref{nLsersic}--\ref{RLsersic} show results obtained 
from fitting Sersic profiles to the full early-type sample, 
as well as to the BCGs.  
Figure~\ref{nLsersic} shows that objects which we identified 
as potentially having two components (like those shown in 
Figure~\ref{twodeVs}) tend to have large values of $n$.  
Figure~\ref{devSersic} shows that, for those objects which we 
flagged as having two components, the luminosities and half-light 
radii derived from the Sersic fits tend to be substantially larger 
than when the fit is restricted to a deVaucouleur profile.  
In some cases the difference in total luminosity is as large as 
one magnitude---this systematic difference dwarfs the 0.02~mag 
uncertainty usually quoted for SDSS photometry.  
Figure~\ref{RLsersic} shows the analog of Figure~\ref{LRe}:  
the $R_e-L$ relation obtained from the Sersic profile fits.
This illustrates that BCGs define a steeper relation than the bulk 
of the population even when parameters derived from Sersic fits are 
used.  

\subsection{The effect of intra-cluster light}\label{testICL}
The previous section showed that BCGs define a steeper size-luminosity 
relation than does the bulk of the early-type galaxy population.  
However, it is known that there is a component of the total luminosity 
of a cluster which is not associated with any particular galaxy in 
it:  this intracluster light (ICL) tends to have a smooth profile 
centered on the cluster center (e.g. Malumuth \& Kirshner 1981; 
Zibetti et al. 2005; Gonzalez et al. 2005; Krick et al. 2006).  
So it is natural to ask if this ICL profile is affecting our estimates 
of the sizes and luminosities of BCGs, thus altering the correlation 
between size and luminosity.  This effect would be particularly severe 
for BCGs which lie at the cluster center.  For such objects, it may be 
more reasonable to treat the surface brightness profile as the sum of 
two components, an inner one which represents the BCG itself, and an 
outer one which represents the ICL.  

To estimate the effect of the ICL on our results, we have used two 
results from the recent literature.  Zibetti et al. (2005) used a 
stacking technique to detect the ICL with high significance in SDSS 
clusters, and concluded that the ICL component has rather low surface 
brightness ($\mu_r>25$~mag~arcsec$^{-2}$):  single SDSS exposures in 
the $r-$band are not sufficiently deep to detect it unambiguously, 
which is why a stacking analysis was required.  
This suggests that contamination from the ICL is unlikely to be a 
strong effect in our case.  
However, recent work by Gonzalez et al. (2005) shows that BCG 
light profiles deviate from a deVaucouleur fit at about the same 
surface brightness values (once the difference between their $I$ 
band and our $r$ band has been accounted for) at which we see 
deviations (c.f. Figure~\ref{twodeVs}).  

To check if the ICL is affecting our results, we made mock 
observations of BCGs of various luminosities as follows.  
For a given BCG luminosity, we require the half-light radius to 
satisfy $R_e\propto L^{0.68}$.  We use this size and luminosity to 
generate a deVaucouleurs surface brightness profile.  
To this profile we add an ICL component, the parameters of which are 
motivated by the study of Zibetti et al..  
Namely, the ICL component is modeled as a second deVaucouleurs 
profile, with the same integrated luminosity as the first component, 
but with half-light radius twenty times larger.  We then fit the 
summed image with a single component deVaucouleurs profile (since 
this is, in effect, what the SDSS pipeline does), and record the 
best-fit values of luminosity and half-light radius.  
Upon repeating this procedure for a range of luminosities, we can 
compare the $R_{\rm BCG+ICL}-L_{\rm BCG+ICL}$ relation with the 
input one.  

In general, the ICL component tends to increase {\em both} the size 
and the total luminosity of the BCG.  Our experiments indicate that 
the net effect of the ICL is to move the BCG along the $R_e-L$ relation.  
Therefore, we find that the ICL component does not lead to a steepening 
of the $R_e-L$ relation.  This conclusion remains unchanged if we use 
Sersic profiles instead, or the particular choice of Petrosian 
quantities used by the SDSS.  (Because the Petrosian quantities do 
not make strong assumptions about the profile shape, one might have 
thought that they were particularly well suited to this comparison.  
However, the Petrosian quantities output by the SDSS pipeline are 
neither seeing-corrected nor have the correct sky subtraction, 
making them less attractive.)

This conclusion also remains unchanged if we assume that the ICL 
component follows a NFW profile (i.e., the profile traced by the 
dark matter in simulations of dissipationless clustering).  For 
NFW concentrations between 5 and 10 (the range associated with 
cluster mass halos), and $L_{\rm ICL}/L_{\rm BCG}$ between 1 and 10, 
the net effect of the ICL is to move the BCG approximately along the 
$R_e-L$ relation.  At larger $L_{\rm ICL}/L_{\rm BCG}$, the ICL 
component results in a flattening rather than a steepening of the 
$R_e-L$ relation.

\section{Discussion}
Whether or not BCGs comprise a special population has been the 
subject of considerable debate for almost fifty years 
(Scott 1957; Sandage 1976; Schechter \& Peebles 1976; 
Tremaine \& Richstone 1977; Bhavsar \& Barrow 1985; 
Oegerle \& Hoessel 1991; Postman \& Lauer 1995; 
Loh \& Strauss 2006), with different authors coming to different 
conclusions.  Semi-analytic galaxy formation models have long 
made the assumption that the central object in a halo is indeed 
special (Croton et al. 2006; Bower et al. 2006).  And recently, 
halo model (see Cooray \& Sheth 2002 for a review) interpretations 
of galaxy clustering strongly suggest that this is indeed the case 
(Berlind et al. 2005; Skibba et al. 2006).  

One of the reasons the subject has received so much attention 
(and controversy!) is that BCGs appear to be quite homogeneous---being 
big and bright, they may be used as standard candles out to large 
distances.  
Their attractiveness as standard candles increases if their nature 
is understood, since the next generation of photometric galaxy surveys 
will identify large numbers of clusters to redshifts of order 0.7.  
This means that large BCG samples will also become available.  
If their nature is understood, then the BCGs themselves may provide 
complementary information about cosmological parameters.  

BCG samples are only just becoming sufficiently large that unambiguous 
conclusions about their nature can be drawn.  The particular question 
we addressed is whether or not BCG formation is dominated by 
dissipation or by dry mergers.  
The dry merger hypothesis seeks to explain why massive early-type 
galaxies, which are expected to have assembled their stellar mass 
recently, host the oldest stellar populations---even when the age 
estimate is luminosity-weighted.  Recent observations suggest that 
dry mergers are a common formation mechanism for massive early-type 
galaxies (e.g., van Dokkum 2005).  Since most BCGs are early-types, 
it is reasonable to ask if they show some signature of unusual 
formation histories.  

Our analysis suggests that the answer to this question is `yes'.  
The optical surface brightness profiles of C4 BCGs appear to be of 
two types---one has the classical deVaucouleur profile 
(Figure~\ref{puredeV}), and the other shows evidence for a second, 
more extended, lower surface brightness component (Figure~\ref{twodeVs}).  
The BCGs which are well-fit by classic deVaucouleur profiles have 
larger half-light radii than normal early-type galaxies of similar 
luminosities (Figure~\ref{LRe}).  The other BCGs also appear to  
follow a steep $R_e-L$ relation.  The low surface brightness 
component in these objects may be due to light from the ICL.  
If so, then the steeper $R_e-L$ relation is also true of the 
BCG components of these objects, since the ICL tends to move 
objects along the $R_e-L$ relation, rather than change its slope 
(Section~\ref{testICL}).  

We also examined the $R_e-L$ relation of the normal early-type 
galaxy population, with a view to seeing if the most luminous 
galaxies tend to have abnormally large sizes whether or not they 
are BCGs.  Although there is clear evidence for curvature in the 
relation defined by the full early-type sample (BCG + non-BCG), 
the evidence for curvature in the early-type sample once BCGs 
had been removed was much less compelling (Figure~\ref{allLRe}).  

The $R_e-L$ relation of the bulk of the early-type galaxy 
population is steeper than the correlation between half mass 
radius and mass of dark matter halos.  
This is consistent with models in which galaxies in low mass halos 
have suffered more dissipation (e.g. Kormendy 1989; Dekel \& Cox 2006).
If massive galaxies have suffered less dissipation than less 
massive galaxies, then our results indicate that BCGs have 
suffered even less dissipation than other massive galaxies of 
similar luminosities.  
Although some of the steepening of the $R_e-L$ relation may 
be due to anisotropic mergers (e.g., Boylan-Kolchin et al. 2006),  
the larger than expected sizes of BCGs also suggest less dissipation, 
so they are consistent with the dry merger formation hypothesis.  
In addition, BCGs show a substantially smaller scatter around the 
Fundamental Plane defined by the bulk of the early-type population 
than does the early-type population itself (Figure~\ref{fpbcg}).  
This might indicate that the formation histories of BCGs involve 
less recent star formation than occurs in early-type galaxies 
of similar mass; below average recent star formation rates would 
be consistent with dry merger formation histories.  

BCGs define the same $g-r$ color-magnitude relation as the bulk 
of the early-type population (Figure~\ref{cmbcg}).  Hence, if 
BCGs form from dry mergers, then BCG progenitors must lie redward 
of this relation (else the product of the merger would lie blueward 
of this relation).  In turn, this suggests that BCG progenitors 
host older stellar populations than is typical for their 
luminosities.  This may provide an interesting constraint on dry 
merger models.  To illustrate why, the Appendix compares our 
measurements with the color-magnitude relation predicted by 
semi-analytic galaxy formation models.  The models suggest that 
BCGs should lie slightly blueward of the relation defined by the 
bulk of the population (Figure~\ref{munichCM})---an effect which 
is not seen in the data (Figure~\ref{cmbcg}).  

The C4 BCGs also define a tight correlation between luminosity and 
dynamical mass $\propto R_e\sigma^2$, although this correlation 
is slightly steeper than it is for the bulk of the early-type 
population (Figure~\ref{MLbcg}).  
This, with the steeper $R_e-L$ relation, implies that they define 
a shallower $\sigma-L$ relation than the bulk of the population 
(Figure~\ref{LSigma}).  These scaling relations are altered even 
for objects which are well described by classical deVaucouleur 
profiles (Figures~\ref{LRe}--\ref{RLsersic}).  
These changes to the scaling relations are consistent with previous 
work based on smaller samples of local BCGs 
(Malumuth \& Kirshner 1981, 1985; Oegerle \& Hoessel 1991).  

The change in slope of the $\sigma-L$ relation may be related to 
studies of curvature in the $M_\bullet-\sigma$ relation.
This is because the most massive galaxies are expected to host 
the most massive black holes.  This expectation is based on the 
strong and tight correlation between $M_\bullet$ and the velocity 
dispersion $\sigma$ of the spheroid/bulge which surrounds it:  
roughly $M_\bullet\propto\sigma^4$ (e.g. Ferrarese \& Merritt 2000; 
Gebhardt et al. 2000; Tremaine et al. 2002).  
$M_\bullet$ also correlates with the luminosity of the bulge, 
although this correlation is not as tight (Magorrian et al. 1995).  
The curvature we see in the $\sigma-L$ relation (Figure~\ref{LSigma}), 
suggests that a single power law may be a better description of the 
$M_\bullet-L_{\rm bulge}$ relation than of $M_\bullet-\sigma$.  
So it is interesting that Ferrarese et al. (2006a,b) find that the mass 
of the central compact object correlates better with bulge mass than 
with $\sigma$ or $L$.   

\acknowledgements
We would like to thank the referee for a helpful report, 
Lucca Ciotti for discussions about dry mergers, and Tod Lauer 
and Mark Postman for discussions about the differences between 
their photometric reductions of local BCGs and those output by 
the SDSS pipeline.  
MB is supported by NASA grant LTSA-NNG06GC19G.  

Funding for the SDSS and SDSS-II
has been provided by the Alfred P. Sloan Foundation, 
the Participating Institutions, the NSF, the US DOE, NASA, 
the Japanese Monbukagakusho, the Max Planck Society 
and the Higher Education Funding Council for England.  
The SDSS website is http://www.sdss.org/.

The SDSS is managed by the Astrophysical Research Consortium (ARC) 
for the Participating Institutions:  The American Museum of Natural 
History, Astrophysical Institute Postdam, the University of Basel, 
Cambridge University, Case Western Reserve University, 
the University of Chicago, Drexel University, Fermilab, 
the Institute for Advanced Study, the Japan Participation 
Group, the Johns Hopkins University, the Joint Institute for 
Nuclear Astrophysics, the Kavli Institute for Particle Astrophysics 
and Cosmology, the Korean Scientist Group, the Chinese Academy 
of Sciences (LAMOST), Los Alamos National Laboratory, 
the Max Planck Institute for Astronomy (MPI-A), 
the Max Planck Institute for Astrophysics (MPA), 
New Mexico State University, the Ohio State University, 
the University of Pittsburgh, the University of Portsmouth, 
Princeton University, the U.S. Naval Observatory, and the 
University of Washington.

\appendix

\begin{figure}
 \centering
 \epsfxsize=\hsize\epsffile{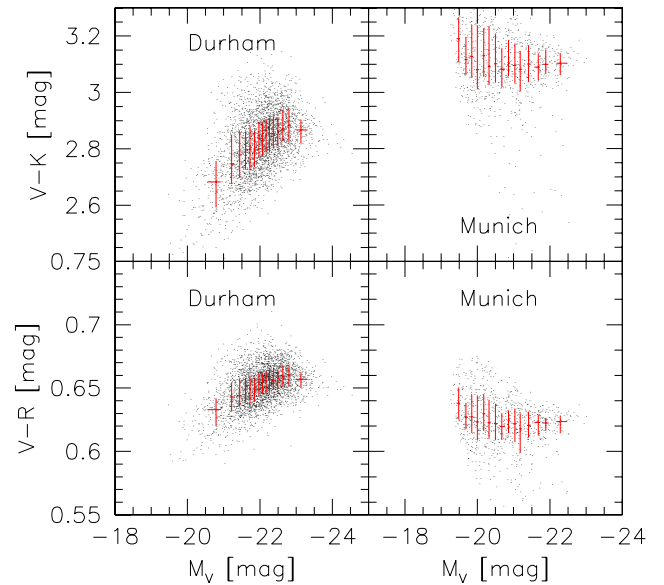}
 \caption{Color-magnitude relations defined by massive galaxies 
          in the Durham and Munich semi-analytic galaxy formation 
          models.  Crosses show the range spanned by the second 
          and third quartiles in color as a function of magnitude
          (the horizontal error bar shows the width of the magnitude 
          bin).}
 \label{durhamMunich}
\end{figure}

\begin{figure}[t]
 \centering
 \epsfxsize=\hsize\epsffile{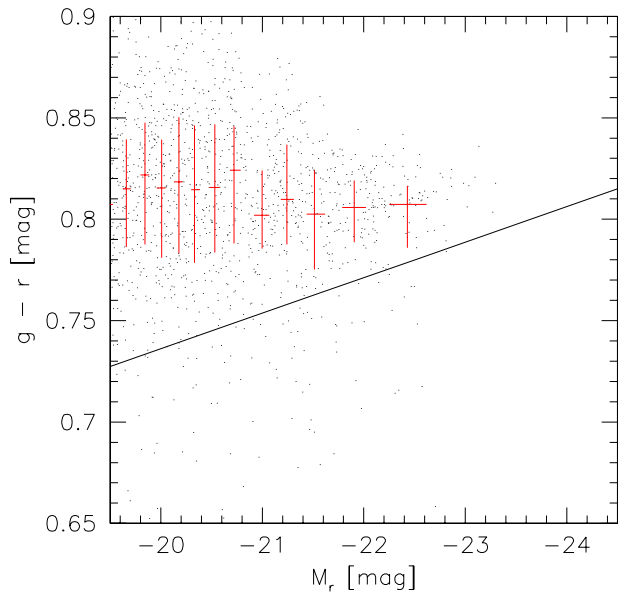}
 \caption{Predicted $g-r$ color-magnitude relation in the 
          Munich model.  Crosses show the range spanned by the second 
          and third quartiles in color as a function of magnitude
          (the horizontal error bar shows the width of the magnitude 
          bin).  Solid line shows the relation in the SDSS.  
          The average color is redder than in the SDSS, and the trend 
          for the most massive objects to be slightly bluer is not 
          seen in the SDSS.  }
 \label{munichCM}
\end{figure}

\section{A: Comparison with semi-analytic galaxy formation models}

The introduction states that semi-analytic galaxy formation models 
make strong assumptions about the formation histories of BCGs.  
This Appendix shows that although the models produce massive 
galaxies which lie on a well-defined color magnitude relation,
different models disagree on the slope and zero-point of this 
relation (Figure~\ref{durhamMunich}).  
Nevertheless, the models do agree that the very most massive 
objects lie slightly blueward of the color-magnitude relation 
defined by the bulk of the bulge-dominated population.  
The SDSS early-type BCGs studied in the main text do not show 
such a blueward trend (Figure~\ref{cmbcg}).  

The `Munich' and `Durham' models we study here were kindly 
made available by Croton et al. (2006) and Bower et al. (2006) 
respectively.  Both are based on the same underlying dark 
matter distribution---that of the Millennium Simulation 
(Springel et al. 2005).  
For the Durham models, we selected BCGs of halos more massive 
than $10^{14}h^{-1}M_\odot$; this resulted in about 3200 objects 
from a $(500h^{-1}$Mpc)$^3$ volume.  The color magnitude relation 
defined by these objects is shown in the left-hand panels of 
Figure~\ref{durhamMunich}.  
Selecting BCGs from the Munich models is less straightforward, 
because information about halo masses is not provided.  
However, the stellar masses of the Durham model BCGs are all 
greater than $1.6\times 10^{10}h^{-1}M_\odot$, and so the right 
hand panels of Figure~\ref{durhamMunich} show the color-magnitude 
relation of the objects which, in the Munich models, have stellar 
masses greater than $2\times 10^{10}h^{-1}M_\odot$.  The 
difference between the Durham and Munich models is striking, 
and is not particularly sensitive to this cut.  This illustrates 
that the semi-analytic modelers have not yet converged on an 
unambiguous prediction for the slope and zero-point of the 
color-magnitude relation.  

For a more direct comparison with the results presented in the 
main text, Figure~\ref{munichCM} shows the predicted $g-r$ color 
as a function of $r$-band magnitude in the Munich models.  
Comparison with Figure~\ref{cmbcg} shows that the predicted 
colors are redder than in the SDSS.  In addition, the trend 
for the most massive objects to be slightly bluer is not seen 
in the SDSS.  If color gradients are responsible for the 
overall offset in color, then they must change along the 
color-magnitude relation to reconcile the predicted bluer 
colors of the most massive objects with the observations.  

\section{B: BCGs and the most luminous galaxies}\label{bigL}

\begin{figure*}[t]
 \centering
 \epsfxsize=0.45\hsize\epsffile{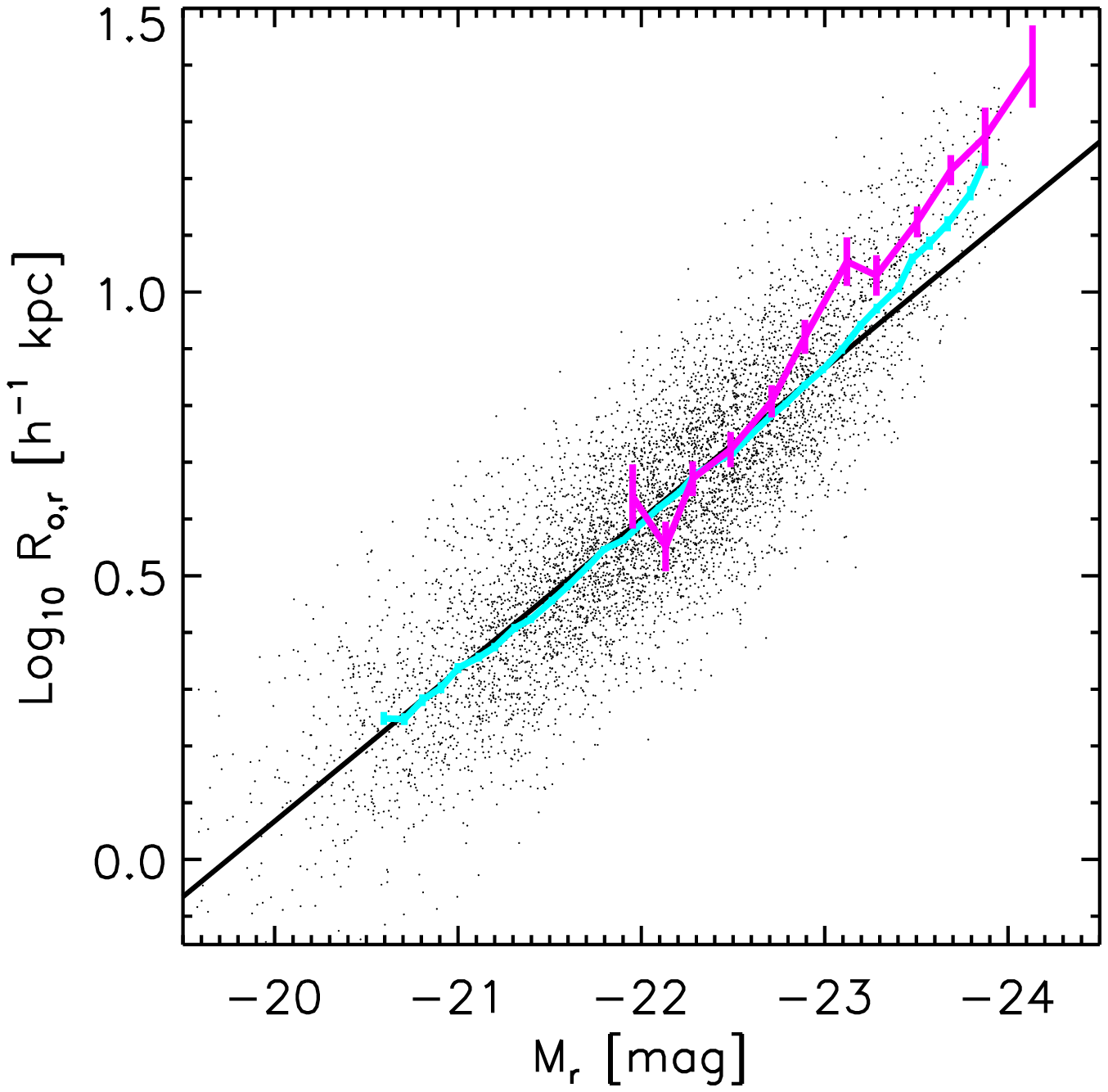}
 \epsfxsize=0.45\hsize\epsffile{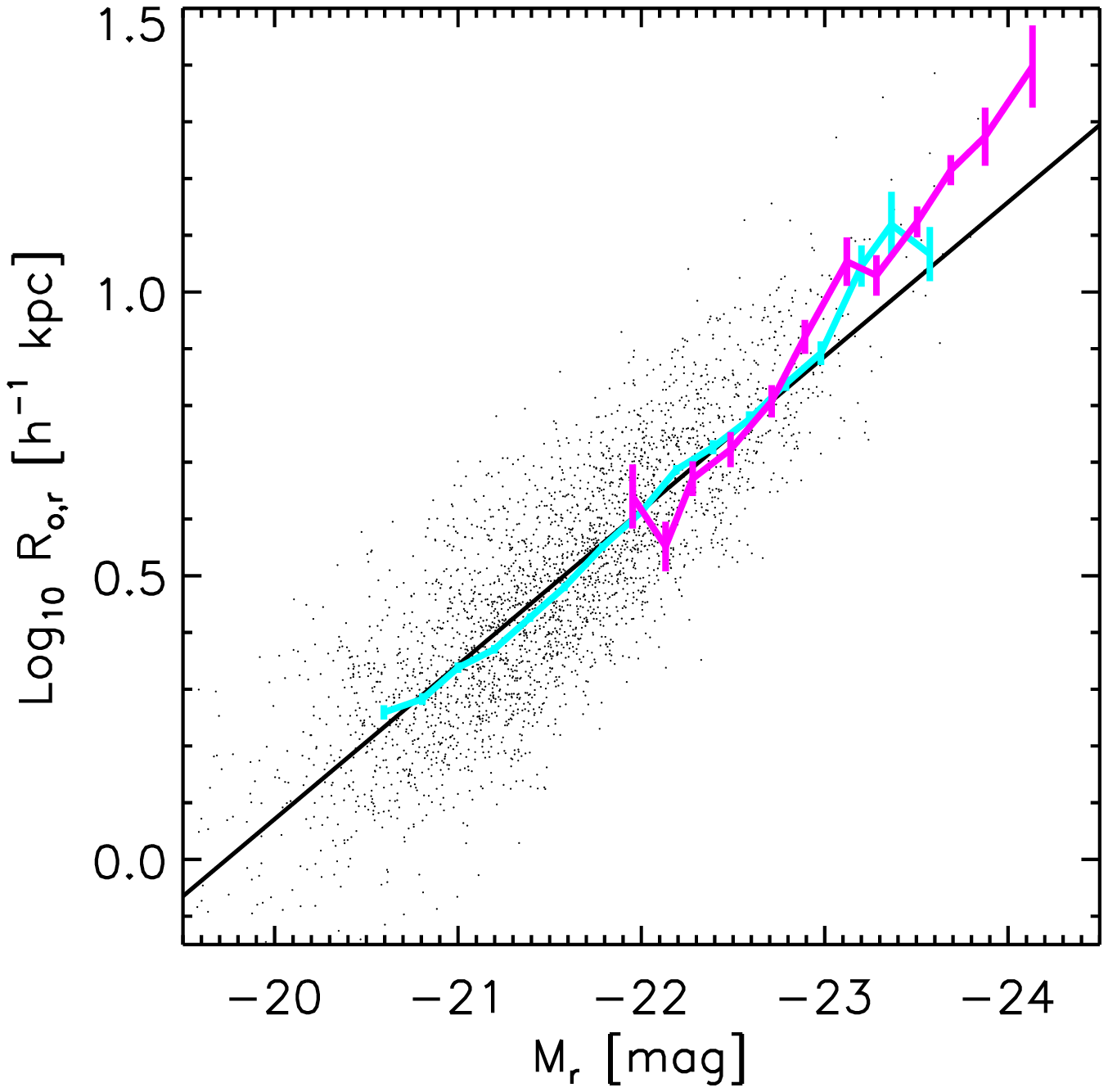}
 \epsfxsize=0.45\hsize\epsffile{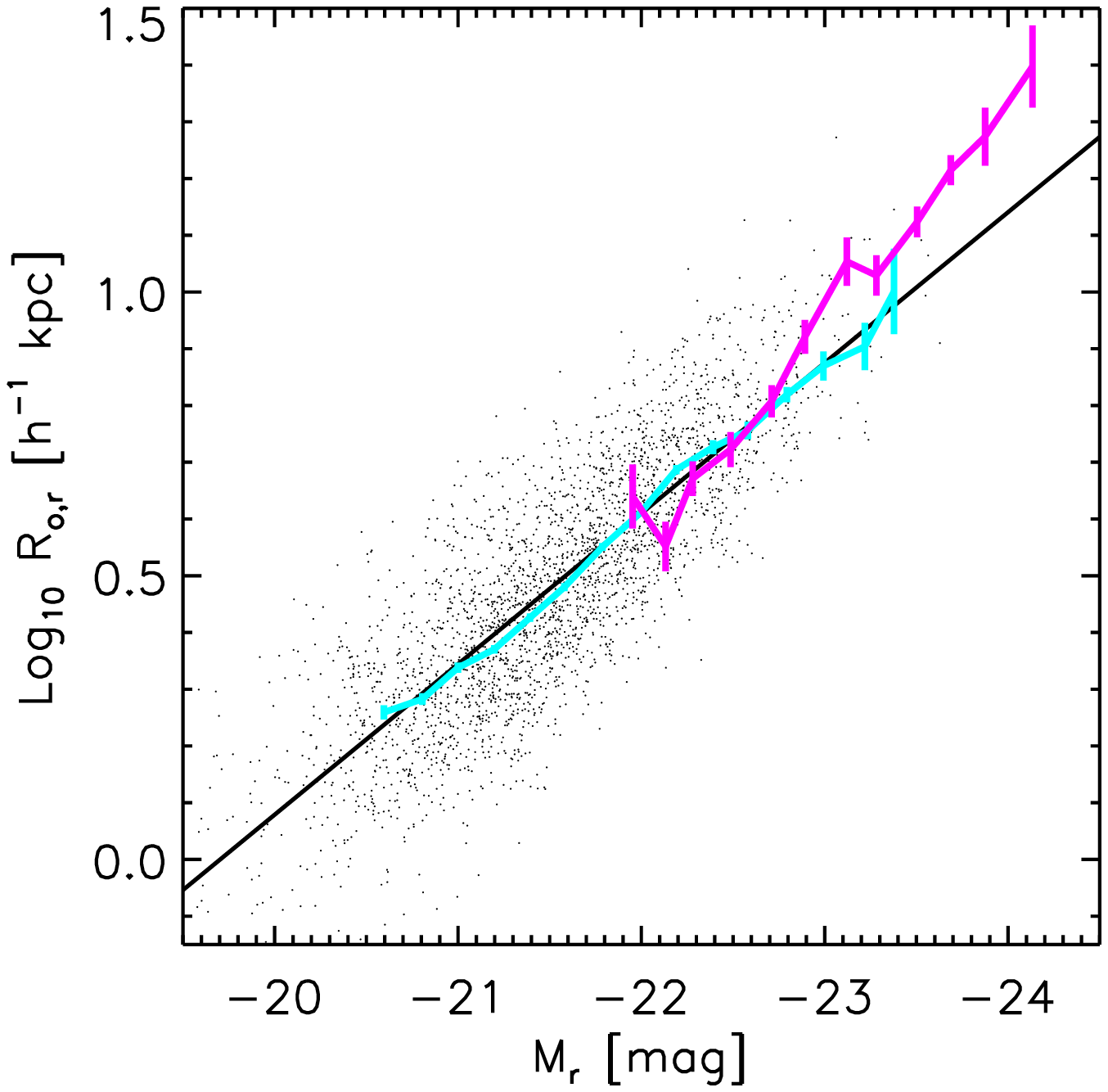}
 \epsfxsize=0.45\hsize\epsffile{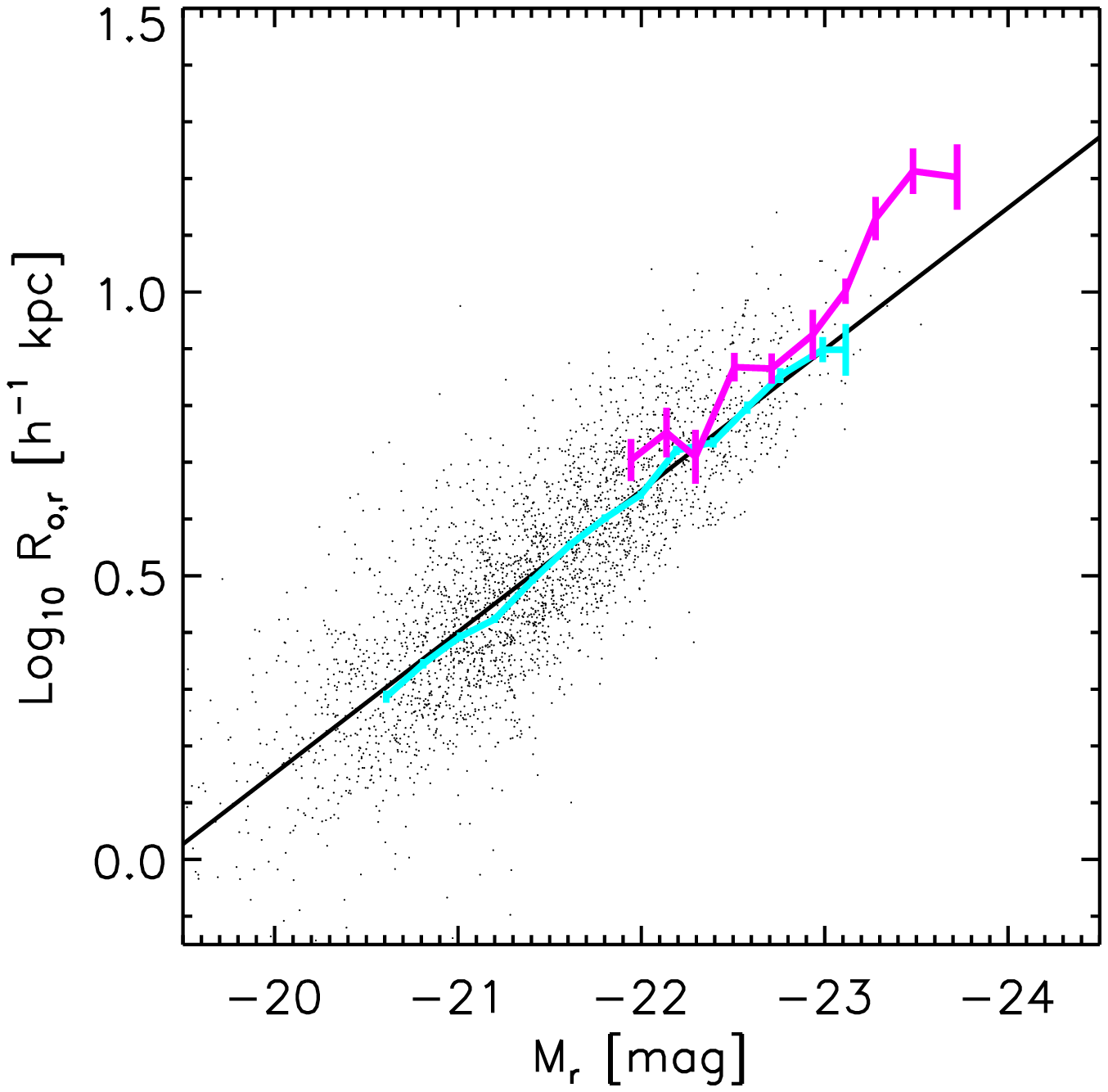}
 \caption{Comparison of the $R_e-L$ relation defined by BCGs and 
          the rest of the early type galaxy population.  
          Magenta jagged solid line (same in all but bottom right panel) 
          shows the median BCG size as a function of luminosity for 
          bins in $L_r$ of width 0.2~mags. 
          Jagged cyan line in each panel shows a similar analysis 
          of the full sample of early types (top left; BCG + non-BCG), 
          of a subsample limited to $z < 0.12$ in which C4 BCGs plus max-BCGs 
          have been removed (top right) 
          and of the same subsample when objects with suspect photometry 
          have also been removed (bottom left).  
          Straight solid line shows a power-law fit to the points 
          in each panel:  it is essentially unchanged (and well 
          described by equation~\ref{lreBCG}) in all but the bottom 
          right panel.  Bottom right panel is similar to top right, 
          but with SDSS parameters instead of those from 
          Hyde et al. (2006).}
 \label{allLRe}
\end{figure*}

So far we have focussed on the fact that BCGs appear to have 
larger than expected sizes for their luminosities.  It is, of 
course, possible that a single power-law is not a good description 
of the $R_e-L$ relation of normal early-types.  Figure~\ref{allLRe} 
presents a study of this correlation for early-types which are not 
BCGs.  The top left panel shows the joint distribution of $R_e$ 
and $L$ in the full Bernardi et al. (2003a) sample, using the photometric 
reductions of Hyde et al. (2006). 
Notice that the lower $R_e$ envelope of the distribution 
appears to curve upward at large $L$, whereas the upper $R_e$ 
envelope is approximately constant.  The solid line shows the 
result of fitting a single power to $\langle R_e|L\rangle$.  
Upper (magenta) and lower (cyan) jagged lines show the median size 
in small bins 
in $L$ in the C4 BCG sample and in the total sample.  At large $L$, 
both samples curve upwards, away from the single power law fit.

\begin{figure*}
 \centering
 \epsfxsize=0.75\hsize\epsffile{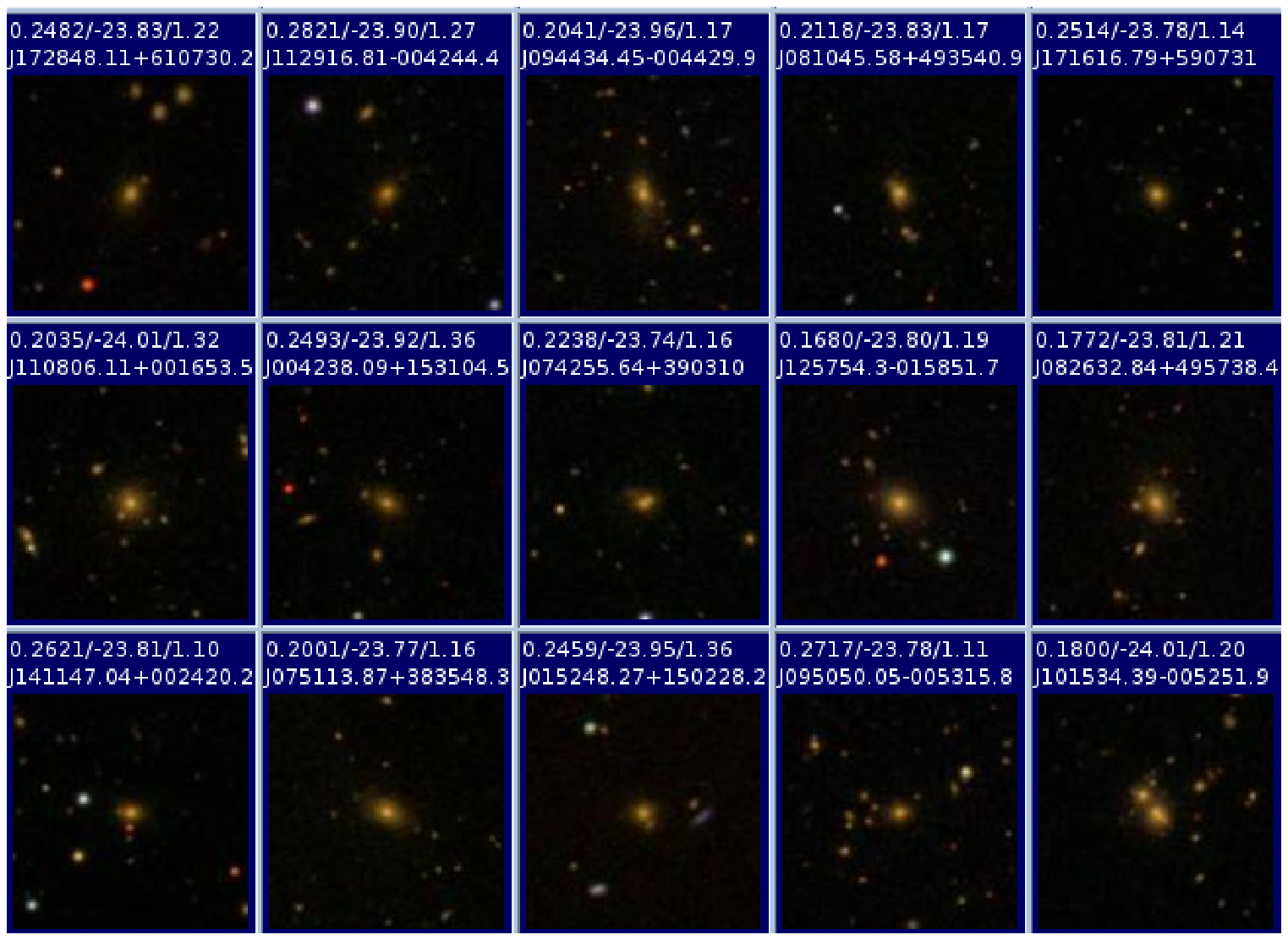}
 \caption{$1.5'\times 1.5'$ fields centered on a random selection of 
          objects with $M_r<-23$ in the SDSS/Bernardi et al. (2003a) sample.  
          The fields tend to be 
          very crowded, with many objects in the field having the same 
          color as the brighter central object, as expected if the 
          central object is a BCG.  Band at top of each image shows 
          the redshift, the absolute magnitude in r-band, $\log_{10}R_e$ 
          of the object and its SDSS ID.}
 \label{bcgM24}
\end{figure*}

The upwards curvature in the full sample (normal plus BCGs) is 
qualitatively consistent with the curvature in the isophotal 
size-luminosity relation for cluster ellipticals recently noted 
by Cypriano et al. (2006).  (In principle, curvature in the 
isophotal size-luminosity relation can result even if there is no 
curvature in the $R_e-L$ relation; we have checked that this 
cannot account for the curvature they report.)  
What causes this curvature?  

\begin{figure*}
 \centering
 \epsfxsize=0.85\hsize\epsffile{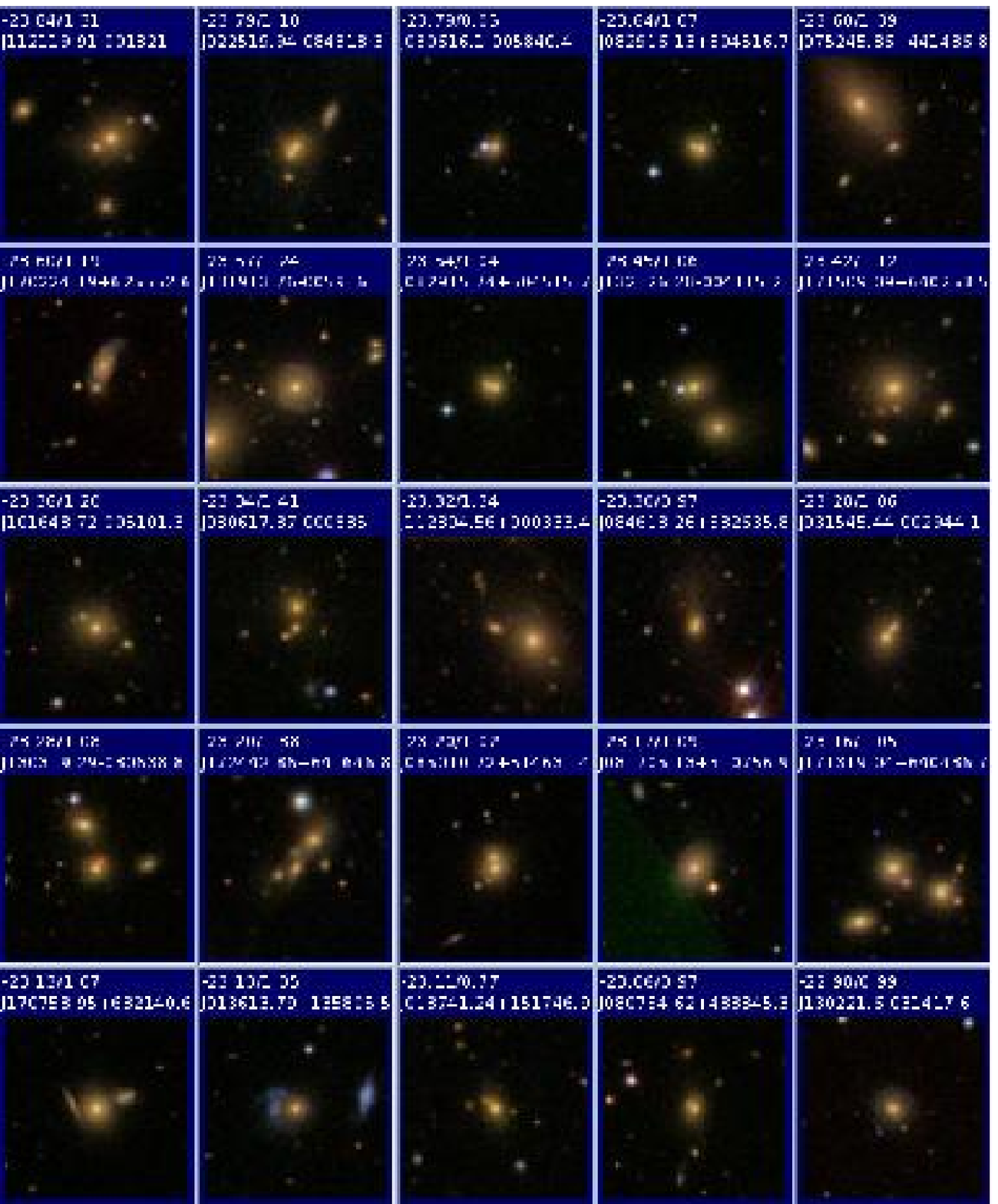}
 \caption{Objects with $M_r<-23$ and $z<0.12$ in the SDSS which are 
          in crowded fields or are contaminated/peculiar galaxies:  
          these objects were present in the 
          non-BCG sample shown in the top right panel of 
          Figure~\ref{allLRe}, but were removed to make the bottom 
          left panel of that Figure.  Images have been ordered by 
          luminosity:  the band at the top of each image gives the 
          absolute magnitude in r-band, $\log_{10}R_e$ and the object ID.}
 \label{throw}
\end{figure*}

\begin{figure*}
 \centering
 \epsfxsize=0.85\hsize\epsffile{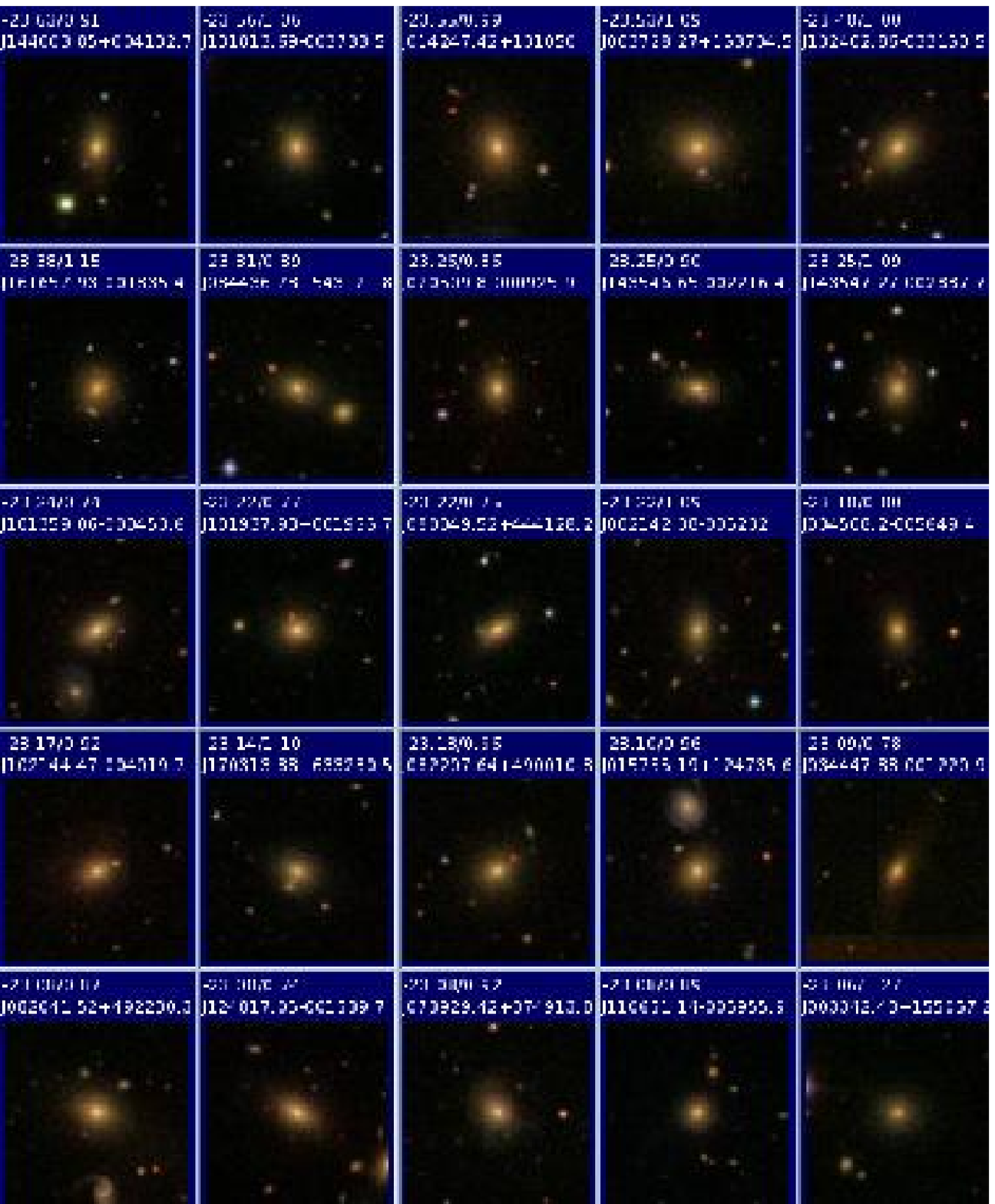}
 \caption{Objects with $M_r<-23$ and $z<0.12$ in the SDSS which are 
          in relatively clean fields or do show strong contamination:  
          these objects make the 
          non-BCG sample in both the top right and bottom left 
          panels of Figure~\ref{allLRe}.  Images have been ordered by 
          luminosity: the band at the top of 
          each image gives the absolute magnitude in r-band, $\log_{10}R_e$ 
          and the object ID.}
 \label{keep}
\end{figure*}

The $R_e-L$ relation is a number-weighted average of the normal 
and BCG relations.  So, if the fraction of objects which are BCGs 
increases with increasing luminosity, then this will produce 
curvature in the same sense as seen in Figure~\ref{allLRe}.  
If all of the curvature is to come from this effect, then the 
fraction of BCGs must increase with luminosity approximately as
 $f_{\rm BCG}(L_r) = [1 + \exp(M_r+23.5)]^{-1}$; 
about half of all objects with $M_r=-23.5$ must be BCGs.  

To illustrate that a substantial fraction of the most luminous 
galaxies are BCGs, Figure~\ref{bcgM24} shows $1.5'\times 1.5'$ 
fields centered on a random selection of objects with $M_r<-23$ 
in the SDSS/Bernardi et al. sample.  
The fields tend to be very crowded, with many objects 
in the field having the same color as the more luminous central 
object, thus strongly suggesting that the central object is a BCG.  

Unfortunately, the vast majority of objects with such luminosities 
in the SDSS lie at redshifts which are substantially larger than 
those over which the C4 catalog is complete---all the objects in 
Figure~\ref{bcgM24} are at $z > 0.2$; the C4 catalog is only 
complete to about $z\sim 0.12$.  This makes it difficult to quantify 
how the BCG fraction increases with luminosity.  
Nevertheless, the top right and bottom left panels in 
Figure~\ref{allLRe} show an attempt to address this issue.  
The points in the top right panel show the result of limiting the 
Bernardi et al. sample to the same redshift range as the C4 catalog 
($z < 0.12$), and then removing the C4 BCGs as well as the objects 
identified as BCGs in the maxBCG catalog of Koester et al. (2007).  
The solid line is a single power-law fit to the points in this panel; 
it is essentially the same as in the previous panel.  The jagged 
magenta line shows the same BCG relation as in the previous panel.  
Although the lower envelope still curves upwards, the upper envelope 
is no longer as constant as before---removing the BCGs has depleted 
this upper envelope substantially.  Nevertheless, the cyan 
jagged line suggests that the objects which survived the BCG cuts do 
still lie above the power-law relation at $M_r<-23$.  
What are these non-BCG objects with abnormally large sizes?  

Recall that the C4 catalog only contains clusters with more than 
ten luminous galaxies, so the early-type sample which remains still 
contains BCGs of less massive clusters and groups.  
Since there is no reason to believe that the formation histories 
of the BCGs of massive clusters are particularly different from 
those of rich groups, it is possible that such objects contribute 
to the curvature in the top right panel of Figure~\ref{allLRe}.  
However, a visual inspection of the images of the objects with 
$M_r<-22.5$ showed that many are actually rather complex.  
Although some have multiple near-neighbours of similar color 
(presumably these are the lower mass groups which are not included 
in the C4 or maxBCG catalogs), others appear to have double nuclei, 
or other unusual features.  Hyde et al.'s estimates of the sizes and 
luminosities do not account for such complexities, making their 
estimates for these quantities suspect.  Therefore, on the basis of 
these images, we classified objects as having suspect photometry or 
not.  Figure~\ref{throw} shows $1.5'\times 1.5'$ fields centered on the 
objects with $M_r<-23$ which we removed, and Figure~\ref{keep} 
shows similar fields centered on the objects which we kept.

The jagged cyan line in the bottom left panel of Figure~\ref{allLRe} 
shows the $R_e-L$ relation of the objects which we felt had more 
reliable photometric reductions.   Notice that the 
difference between the jagged and solid black lines, which was 
obvious in the top right panel, has gone away, even though the 
actual values of $R_e$ and $L$ played no role in determining which 
objects were kept.  

As a check, the bottom right panel of Figure~\ref{allLRe} shows 
the SDSS reductions of the same objects shown in the top right 
panel.  Recall that these reductions slightly underestimate the 
luminosities and sizes of bright objects in crowded fields 
(such as BCGs); the solid line which shows a fit to these points 
has a shallower slope than in the other panels.  The BCGs 
clearly lie above this relation at large $L$, whereas the non-BCGs 
do not.

Thus, although our results do not exclude the possiblity that 
there is real curvature in the $R_e-L$ relation of normal early-type 
galaxies, they do suggest that to estimate this curvature 
reliably, one must be careful to ensure that the sample contains 
no BCGs or contaminated galaxies.  In principle, one can use recent 
halo model interpretations (see Cooray \& Sheth 2002 for a review) 
of the luminosity dependence of galaxy clustering to estimate this 
effect, but this is beyond the scope of the present work.

\end{document}